\documentclass[aps,prd,preprint,showpacs,floatfix,preprintnumbers,nofootinbib,superscriptaddress,showkeys]{revtex4}
\usepackage[utf8]{inputenc}
\usepackage{color}
\usepackage{fancybox}
\usepackage{hhline}
\usepackage{dcolumn}
\usepackage{textcomp}
\usepackage{epsfig,graphics,graphicx}
\usepackage{amsfonts,amssymb,amsmath}
\usepackage{pifont}
\usepackage{bm}
\usepackage{longtable} 
\usepackage{appendix}
\usepackage{lscape}
\usepackage[mathscr]{euscript}
\usepackage{mathrsfs}
\usepackage{multirow}
\usepackage{rotating}
\usepackage{color}   
\usepackage[dvipsnames]{xcolor}

\newcommand{\beq}{\begin{equation}}
\newcommand{\eeq}{\end{equation}}
\newcommand{\bea}{\begin{eqnarray}}
\newcommand{\eea}{\end{eqnarray}}

\newcommand{\eps}{\epsilon}
\newcommand{\ord}[1]{{\cal{O}}\left( #1 \right)}
\newcommand{\B}{{\bf B}}
\newcommand{\Bdag}{{\bf B^\dagger}}
\newcommand{\p}{{\mathfrak{p}}^0}

\DeclareFontFamily{OT1}{pzc}{}
\DeclareFontShape{OT1}{pzc}{m}{it}%
              {<-> s * [0.900] pzcmi7t}{}
\DeclareMathAlphabet{\mathpzc}{OT1}{pzc}%
                                 {m}{it}
\DeclareMathAlphabet{\mathcalligra}{T1}{calligra}{m}{n}
\usepackage{hyperref}
\hypersetup{pdfauthor={me}, colorlinks=true, citecolor=blue, urlcolor=blue, linkcolor=black}
\begin{document}
\preprint{\vbox{\hbox{ JLAB-THY-14-1866} }}
\title{\phantom{x}
\vspace{-0.5cm}     }
\title{The $N_c$ dependencies of baryon masses: Analysis with \\ Lattice QCD and Effective Theory}
\author{A.~Calle~Cord\'on}\email{alvaro.calle@um.es}
\affiliation{Thomas Jefferson National Accelerator Facility, Newport News, VA 23606, USA.}
\affiliation{Departamento de F\'isica, Universidad de Murcia, Murcia, E-30071 Murcia, Spain}
\author{T.~DeGrand}\email{thomas.degrand@colorado.edu}
\affiliation{Department of Physics, University of Colorado, Boulder, CO 80309, USA}
\author{J.~L.~Goity}\email{goity@jlab.org}
\affiliation{Thomas Jefferson National Accelerator Facility, Newport News, VA 23606, USA.}
\affiliation{Department of Physics, Hampton University, Hampton, VA 23668, USA.}
%
 
\begin{abstract} Baryon masses at varying values of $N_c$ and light quark masses are studied with Lattice QCD and the results are analyzed in a low energy effective theory based on  a combined framework of the  $1/N_c$ and Heavy Baryon Chiral Perturbation Theory expansions.  Lattice QCD results for $N_c=3$, 5 and 7 obtained in  quenched calculations, as well as results for unquenched calculations for $N_c=3$, are used for the analysis. The results are consistent with a previous analysis of $N_c=3$  Lattice QCD results,   and in addition permit the determination of  sub-leading in $1/N_c$ effects  in the spin-flavor singlet component of the baryon  masses as well as in the hyperfine splittings. 
 \end{abstract}

\pacs{11.15-Pg, 11.30-Rd, 12.39-Fe, 14.20-Dh}
\keywords{Baryons, Lattice QCD,  Large N, Chiral Perturbation Theory}

\maketitle

\tableofcontents

\section{Introduction}\label{sec:Intro} 

 Lattice QCD (LQCD) calculations at varying number of colors $N_c$  provide the opportunity  for understanding more quantitatively important aspects of hadronic physics  which have been  qualitatively  described in terms of  order of magnitude estimates based on $N_c$ power countings.       %
Until recently,  LQCD analyses at  $N_c>3$   had  been carried out only for pure gluon dynamics and for mesons, where first studies addressed confinement in $SU(2)$, $SU(3)$ and $SU(4)$ Yang-Mills gauge theories~\cite{Teper:1997tq},   tests of the 't Hooft scaling
in  $g^2 N_c = \lambda$,
the relation between the QCD scale, $\Lambda_{QCD}$ and the gauge coupling $g$ for different $N_c$ 
in~\cite{Lucini:2001ej,DelDebbio:2001sj,Allton:2008ty}, calculations of the glueball spectrum~\cite{Lucini:2004my,Lucini:2010nv}, meson spectroscopy  ~\cite{DelDebbio:2007wk,Bali:2008an,Bali:2013kia}, masses and decay constants and their scalings with $N_c$ ~\cite{Narayanan:2005gh,Hietanen:2009tu,DeGrand:2012hd}, the topological susceptibility~\cite{Vicari:2008jw}.
 For a review of these developments the reader can consult   Ref.~\cite{Lucini:2012gg}.

Recently, the extension to baryons, which is the focus of the present work, was performed by one of us, T. DeGrand in Ref.~\cite{DeGrand:2012hd}. 
Baryon masses for $N_c=3,$ 5 and 7 were studied
in the quenched approximation and at  pion masses above 400 MeV. The results give a striking confirmation of  the 
large $N_c$ QCD  predictions for baryons, namely the  $N_c$ scalings  of   baryon masses and of  hyperfine (HF) mass
 splittings with $N_c$. That work also provides  information on the  quark mass dependence of baryon masses, which is exploited
 in the present work for the purpose of  understanding the effective low energy theory based on the 
combined $1/N_c$ expansion and Heavy Baryon Chiral Perturbation Theory 
(HBChPT)~\cite{Jenkins:1995gc,FloresMendieta:1998ii,FloresMendieta:2006ei,CalleCordon:2012xz,Cordon:2012nm,Cordon:2013era}. 

The importance of the $1/N_c$ expansion in BChPT was pointed out long ago with the observation that the inclusion of the  spin $\frac{3}{2}$ baryons leads to important cancellations in chiral loop corrections to axial currents \cite{Jenkins:1990jv,Jenkins:1991ne}. It was realized that such cancellations are a result of the constraints on the meson-baryon couplings in large $N_c$ \cite{Dashen:1993ac,Dashen:1993as}, which gives  rise to a contracted dynamical spin-flavor symmetry in baryons, which in particular requires the inclusion of the higher spin baryons, up to spin $N_c/2$. This leads to the implementation of  a baryon chiral Lagrangian following the strictures of the $1/N_c$ expansion \cite{Jenkins:1995gc}, a framework  that has been further developed and applied in \cite{FloresMendieta:1998ii,FloresMendieta:2006ei,CalleCordon:2012xz,Cordon:2012nm,Cordon:2013era,FloresMendieta:2012dn}. The virtue of such an effective field theory (EFT) is that it incorporates the dictates of the spin-flavor symmetry, emerging in the baryon sector of QCD at $N_c\to \infty$, into the chiral expansion and allowing for an unambiguous treatment of the spin-$\frac{3}{2}$ resonance as an explicit degree of freedom in the EFT.

In recent works~\cite{CalleCordon:2012xz,Cordon:2012nm,Cordon:2013era}, the EFT  combining  the $1/N_c$  expansion and HBChPT was studied and applied to non-strange baryon masses and axial couplings (for a recent study of axial currents in $SU(3)$ see Ref. \cite{FloresMendieta:2012dn}).  LQCD results at $N_c = 3$ were used to determine low-energy constants (LECs), and to test the low energy expansion.  At  small  enough quark masses,  where the low energy expansion holds, the $1/N_c$ expansion is encoded in LEC's.  Each operator in the chiral Lagrangian carries a LEC, which can be expanded in powers of $1/N_c$. In order to determine the LECs, it  is therefore necessary to have information at different values of $N_c$.
This point is investigated with the LQCD results in the present work.

As has been already shown in Refs.~\cite{DeGrand:2012hd,DeGrand:2013nna}, LQCD baryon masses reproduce remarkably well the rotor spectrum predicted by the $1/N_c$ expansion at large $N_c$. The most general mass formula is:
\beq
m_\B(S,N_c,M_\pi)=N_c \, {\bf m_0}(N_c,\,M_\pi)+\frac{{\bf C_{HF}}(N_c,\,M_\pi)}{N_c}\;S(S+1) +\ord{\frac{(S(S+1))^2}{N_c^2}}\, ,
\label{eq:rotor}
\eeq
where $\bf m_0$ and $\bf C_{HF}$ start at $\ord{N_c^0 M_\pi^0}$,  and  can be determined  with LQCD calculations and also studied with the EFT. Using the LQCD results \cite{DeGrand:2012hd}, one can obtain an estimate: 
\bea
{\bf C_{HF}} &\sim&\frac{4 N_c}{N_c(N_c+2)-3}\left(m_\B(S=\frac{N_c}{2})-m_\B(S=\frac{1}{2})\right) \, ,\nonumber\\
{\bf m_0}    &\sim&\frac{1}{N_c}\left(\overline{m}_\B-\left(\frac{1}{3}+\frac{1}{N_c}+\ord{ N_c^{-3}}\right)\left(m_\B(S=\frac{N_c}{2})-m_\B(S=\frac{1}{2})\right)\right)\, ,
\eea
where $\overline{m}_\B$ is the average mass (or center of gravity) of the spin-flavor multiplet.
A test of the rotor spectrum is provided by the ratio of the HF splittings  between the  $S=\frac{3}{2},~\frac{1}{2}$ states, and the $S=\frac{N_c}{2},~(\frac{N_c}{2}-1)$ states. The above formula gives $1.66$ for $N_c=5$, while for all values of $M_\pi$ LQCD gives $1.5$ to $1.6$. For $N_c=7$ the formula gives $2.33$ while LQCD gives $2$ to $2.2$. These results are  remarkably close to what is expected in the large $N_c$ limit.

 There is one important point to mention concerning the definition of the $1/N_c$ expansion. Since the $1/N_c$  expansion compares different theories, there is an ambiguity in the very definition of the expansion. The ambiguity stems from the fact that the theories are defined at  a renormalization scale, and that quantities run differently (by subleading corrections in $1/N_c$) in the different theories.  At the level of the fundamental parameters of QCD, the expansion can be defined by  requiring  that the 't Hooft coupling and the quark masses are the same, i.e. independent of $N_c$,  at a  chosen renormalization scale $\mu_0$. A different possibility is to choose hadronic quantities or  observables which are  $\ord{N_c^0}$, such as $M_\pi$ and $F_{\pi}/\sqrt{N_c}$,  and require them to be independent of $N_c$.  What is precisely meant by the $1/N_c$ expansion will therefore depend on the choice  of definition scheme.  In the present work a scheme defined at hadronic level and very similar to the one just mentioned will be utilized.

While performing the present study, it became apparent that the quenched LQCD data alone is not quite sufficient for establishing the robustness of the analysis, and thus it becomes necessary to include also results with dynamical quarks at $N_c=3$. This allows for a better control of the low quark mass domain. A careful discussion of the issues and viability of combining quenched and unquenched results will be presented.

 This work is organized as follows:
  Section \ref{sec:BChPT} gives a brief overview of the combined HBChPT and $1/N_c$ expansion.
  Section  \ref{sec:LQCD}  contains the LQCD results at varying $N_c$.
    Section \ref{sec:Analysis}   contains the analysis, and Section \ref{sec:Conclusions} gives the conclusions. One Appendix gives explicit expressions for the finite parts of self energies and wave function renormalization factors needed in the calculations.

\section{Baryon masses in the combined $1/N_c$ and chiral expansions}
\label{sec:BChPT}


Baryon observables  have peculiar scalings in $N_c$: their masses scale as $N_c$, while the mass differences between states with spins $\ord{N_c^0}$ scale as $1/N_c$. The pion couplings to baryons scale as $\sqrt{N_c}$, which has profound consequences for the baryon chiral expansion. In fact, these peculiar scalings give rise in the large $N_c$ limit to a dynamical spin-flavor symmetry for baryons.  For $N_f$ degenerate flavors, that dynamical symmetry is a contracted  $SU(2N_f)$ symmetry   \cite{Gervais:1983wq,Gervais:1984rc,Dashen:1993as,Dashen:1993ac}, which is broken by the fact that $N_c$ is finite. For $N_c$ sufficiently large those effects can be expanded in powers of $1/N_c$. Although there is no rigorous proof, there are phenomenological indications as well as the LQCD results analyzed here  which  suggest that an expansion, modulo the subtleties discussed below, can be implemented  for $N_c$ down to the real world's value $N_c=3$.

 When the $1/N_c$ expansion is combined with the low energy expansion, it is found that the chiral and $1/N_c$ expansions do not commute due to the presence  of the ratio $M_\pi/(m_\Delta-m_N)=\ord{p N_c}$ in the non-analytic pieces of chiral loop contributions \cite{Cohen:1996zz}. In the meson sector,   the expansions do commute with each other \cite{Kaiser:2000gs} except for the effects of  the $\eta-\eta'$ mixing,   which involve the product $N_c (m_s-\hat{m})$.   ($\hat{m}$ is the average nonstrange quark mass.)
In particular, the non-commutativity of the expansions requires that the two power countings ought to be linked for a definite EFT to be defined. A linking which seems to be the most adequate for the real world baryons  is the $\xi$-expansion \cite{CalleCordon:2012xz},
where the countings  are linked according to $\ord{1/N_c}=\ord{p}=\ord{\xi}$. The chosen power counting scheme determines the  terms in the effective chiral Lagrangian at each given order of the expansion.  
In the $\xi$ expansion, the Lagrangian for the combined BChPT and $1/N_c$ expansions to order $\xi$, following Refs. \cite{Jenkins:1995gc,CalleCordon:2012xz}, reads:
 \bea
{ \cal{L}}_\B^{(1)}&=&\Bdag\left(i D_0 +  \mathring{g}_A
 u^{ia}G^{ia}-\frac{m_{2}}{N_c}-\frac{C_{HF}}{N_c}{ \vec{S}^2}-\frac{c_1}{2} N_c\; \chi_+\right)\B,
\label{eq:Lagrangian-LO}
 \eea
where only the case of two flavors is  considered.  $\B$ is the symmetric spin-flavor baryon multiplet with states $S=I=\frac{1}{2},\cdots,\frac{N_c}{2}$; $G^{ia}$ are the spin-flavor generators of $SU(4)$ with $i$ spatial indices and $a$ isospin indices and matrix elements $\ord{N_c}$. The pions  reside in the chiral covariant  derivative $D_0$ and in $u^{ia}=\frac{1}{2}\,Tr(\tau^a u^i)$, where $u^i=-\frac{1}{F_\pi} \partial^i \vec{\pi}\cdot\vec{\tau}+\cdots$, and the quark masses reside in $\chi_+=2 M_\pi^2+\cdots$. The LECs  $m_2$,  $ \mathring{g}_A$, $C_{HF}$, and $c_1$ are $\ord{N_c^0}$. As defined here, the axial coupling $ \mathring{g}_A$ is related  to the one of the nucleon at $N_c=3$  by $ \mathring{g}_A=\frac{6}{5}\, g_A$, where $g_A=1.27$ is the well known nucleon axial coupling.

The baryon masses  to one loop are as follows  \cite{CalleCordon:2012xz}:
\bea
m_\B(S) &=& N_c \, m_0(N_c)  + c_1\;N_c M_\pi^2  + \frac{C_{HF}}{N_c} S(S+1) \nonumber\\
& + &
\langle \B \vert\;\frac{
(\delta\Sigma^{UV~finite}_{(1-loop)}+\delta\Sigma^{CT})\arrowvert_{\p=0}}{1-\delta Z^{UV~finite}_{(1-loop)}-\delta Z^{CT}}
\vert \B\rangle\; ,
\label{eq:baryon-mass}
\eea
where $S$ is the spin of the baryon,   and  $m_0(N_c)=m_0+\frac{m_{1}}{N_c}+\cdots$. The rest mass which is  removed by the heavy baryon expansion is the term $m_0+\frac{m_{1}}{N_c}$. The one-loop finite pieces are evaluated  in dimensional regularization and MS scheme at a renormalization scale $\mu$, whose dependence is canceled by that of the counter-terms (indicated with the label $CT$). $\p$ is the residual baryon energy, and as usual, the correction to the wave function renormalization factor is defined by $ \delta Z=\partial \delta\Sigma/\partial\p$. The last term on the RHS of Eq.~\eqref{eq:baryon-mass} contains the contributions $\ord{\xi^2}$ and $\ord{\xi^3}$.
 All the details on the derivation of Eq.~\eqref{eq:baryon-mass}  are found in Ref.~\cite{CalleCordon:2012xz}.
 
 The UV divergences are  given  by \cite{CalleCordon:2012xz}: 
 \bea
 \label{eq:dSigmaUV}
 \delta\Sigma(S)^{UV}\arrowvert_{\p=0}&=&\frac{ \mathring{g}_A^2 C_{HF}}{16 \pi^2 \mathring F_0^2}\left(1+\frac{4}{N_c}\right)\left(-\frac{9}{8} M_\pi^2+\frac{C_{HF}^2}{N_c^2}(3+5\, S(S+1))\right)\lambda_\epsilon\nonumber\\
  \delta Z(S)^{UV}&=&\frac{3 \mathring{g}_A^2 }{512\, \pi^2 \mathring F_0^2}(N_c+4)\left(3 M_\pi^2-\frac{8\,C_{HF}^2}{N_c^2}(3+2 S(S+1))\right)\lambda_\epsilon\;,
 \eea
 where $\mathring F_0\equiv\mathring{F}_\pi\sqrt{3/N_c}$ with $\mathring F_\pi$ the pion decay constant in the chiral limit,   $\lambda_\epsilon=\frac{1}{\eps}-\gamma+\log 4\pi$, and
the counter-terms    necessary for the renormalization of the self-energy at $\ord{\xi^3}$ read as follows \cite{CalleCordon:2012xz}:
\bea
\label{eq:dSigmaCT}
\delta \Sigma^{CT}(\p = 0) (S) &=&
 \mu_1  M_\pi^2 +\left (\frac{\mu_2}{N_c}  M_\pi^2+
\frac{C_{HF1} }{N_c^2}\right)S(S+1)  +
\frac{m_3 }{N_c^2}  
+
\frac{m_4 }{N_c^3}  
+ \frac{C_{HF2}}{N_c^3}(S(S+1))^2 , \nonumber\\
\label{eq:dZCT}
\delta Z^{CT}(S) &=& 
 z_1\, N_c\, M_\pi^2  + 
\frac{w_1 }{N_c} 
+ \frac{w_2 }{N_c}S(S+1) +\ord{\xi^2} .
\eea
To one loop,   $\delta \Sigma^{CT}$ has terms   $\ord{\xi^2}$ and  $\ord{\xi^3}$, and $\delta Z^{CT}$ has terms   $\ord{\xi}$ and  $\ord{\xi^2}$. In the result for the masses, Eq. (\ref{eq:baryon-mass}), only the terms   $\ord{\xi}$ in  $\delta Z^{CT}$ are relevant.  
The LECs depend on $\mu$ in order to render the result for the masses $\mu$ independent.  Comparison  with Eq. (\ref{eq:dSigmaUV})  shows that several of the counter-terms are finite, as they are not required to cancel the UV divergences shown in Eq. (\ref{eq:dSigmaUV}).
The masses so calculated are accurate to $\ord{\xi^3}$. For completeness, the explicit expressions of the finite pieces of the self-energy are given in the Appendix.  

In the large $N_c$ limit,   the spin-flavor singlet one-loop contributions to the baryon self-energy  show UV finite  terms $\ord{M_\pi^3 N_c}$, and both UV   finite and  divergent terms $\ord{\p M_\pi^2 N_c}$ which affect the wave function renormalization factor  \cite{CalleCordon:2012xz} .    This implies the breakdown of the low energy expansion for the spin-flavor singlet components of the self-energy  in the strict  limit $N_c\to \infty$.  On the other hand,  the spin-flavor non-singlet components of the self-energy,  i.~e., the  HF  splittings, are suppressed by at least one power of $1/N_c$.  Cancellations of contributions that violate that power behavior lead to an improved convergence of the  low energy expansion of the HF splittings. This is shown in the analysis below, where  HF splittings are consistently described in a larger range of $M_\pi$ than the  spin-flavor singlet masses.

\begin{center}
\begin{figure}[h!!!!]
\centerline{\includegraphics[width=9.5cm,angle=-0]{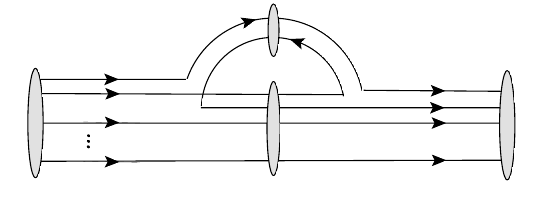}}
\caption{Dominant contribution by pion-loop to baryon self-energy ($\ord{N_c}$). Diagrams with a quark loop  are suppressed   by a factor $1/N_c$. The blobs indicate hadrons.
}
\label{fig:PionLoop}
\end{figure}
\end{center}

It is important to emphasize that the effects of quenching do not qualitatively change the above arguments. In fact, for each operator which appears in the effective Lagrangian, the leading LEC of $\ord{N_c^0}$ should be the same in quenched and unquenched cases. At the quark level one can visualize  the leading contributions in $N_c$ to masses and wave function renormalization factors  by  the diagram  shown in  Fig. \ref{fig:PionLoop}, which corresponds to the quenched case.

Quantitatively, quenching does have an effect, which was first described by
Labrenz and Sharpe~\cite{Labrenz:1996jy,Sharpe:1997by}: In ``real'' QCD with dynamical sea
quarks, the generic baryon mass is
\beq
M_B=M_0 + c_2 M_\pi^2 + c_3 M_\pi^3 + C_{4L}M_\pi^4 \log M_\pi + c_4 M_\pi^4+\dots
\eeq
(the $c_3$ term arises from the one loop pion exchange graph). In quenched QCD,
the mass is
\beq
M_B^q=M_0^q + [\delta c_1^q M_\pi + \delta c_{2L}^q M_\pi^2\log M_\pi]
   + c_3^q M_\pi^3 + C_{4L}^qM_\pi^4 \log M_\pi + c_4^q M_\pi^4+\dots
\eeq
The superscript $q$ reminds us that the quenched coefficients need have no connection to
the unquenched ones.
The extra terms are the quenching artifacts. $\delta=m_0^2/(48\pi^2 f^2)$ and $m_0$ is a number
 associated with the annihilation graph $q \bar q\rightarrow q \bar q$.
The extra terms come when the baryon emits and absorbs $q \bar q$ pairs which would like to iterate into the
 eta (and
then not be present in a chiral expansion), but because it is quenched approximation, the iteration
terminates. It gives a double pole chiral loop.
For the nucleon, $c_1^q= 3\pi/2(3F-D)^2$. As we are interested in large $N_c$, a later paper by
Chow and Rey~\cite{Chow:1997cu} makes a useful point:
the $m_0^2$ parameter in $\delta$ is also $O(1/N_c)$, so $\delta$ scales like $1/N_c^2$.  This means that
 both of these quenching artifacts, the extra term in the expansion and the difference between
quenched and unquenched LEC's, are nonleading in $1/N_c$ compared to the usual chiral factors.
They would appear as  one of many corrections to the leading terms in the chiral expansion.

One could opt to restrict the analysis of this work  only to the quenched data sets of Ref.~\cite{DeGrand:2012hd}. However, 
  the spectroscopies of  the  $SU(3) $  quenched data set  and of the unquenched $SU(3)$  data sets, conveniently scaled by using the chiral limit $F_\pi$ to set the scale,  are found to  be  consistent within uncertainties and so  phenomenologically can be combined.  A test
  using  only quenched or unquenched data will also be done, in order to  avoid
combining quenched and unquenched data sets into a single fit.

\section{Lattice data sets}  \label{sec:LQCD}

The lattice data for $N_c=5$ and $7$ is based largely on that of Ref.~\cite{DeGrand:2012hd}.
Readers should refer there,  and to a later paper focused on flavor $SU(3)$ symmetry,~\cite{DeGrand:2013nna}, for more details. Here is a brief summary:
Simulations use the usual Wilson plaquette gauge action, with clover fermions
 with normalized hypercubic smeared links
as their gauge connections\cite{Hasenfratz:2007rf}.
The clover coefficient is fixed at its tree level value,
$c_{SW}=1$. The code is a version of the publicly available package of the MILC collaboration~\cite{MILC}.

All simulations are performed in quenched approximation.
 The simulation volumes were all $16^3\times 32$ sites.
The bare gauge couplings were (roughly) matched so that pure gauge observables were the same on all three
$N_c$'s,
so as to match discretization and finite volume effects.
The observable chosen to do the comparison was
the shorter version of the Sommer parameter \cite{Sommer:1993ce}
$r_1$, defined in terms of the force $F(r)$ between static quarks,
$r^2 F(r)= -1.0$ at $r=r_1$. The real-world value is $r_1= 0.31$ fm \cite{Bazavov:2009bb},
and with it the common lattice spacing is about 0.08 fm.

Simulation parameters are reported in Table \ref{tab:par}. The masses of the pseudoscalar and vector mesons,
the pseudoscalar decay constant, and the baryon masses are shown in Tables \ref{tab:su5}-\ref{tab:su7}, and the HF splittings are shown in Tables \ref{tab:su3HF} through \ref{tab:su7HF}.

\begin{table}
\begin{tabular}{c c c c}
\hline
                 & $SU(3)$   & $SU(5)$   &   $SU(7)$ \\
\hline
$\beta$          & 6.0175  & 17.5 & 34.9 \\
configurations   &  80     & 120   & 160   \\
$r_1/a$          & 3.90(3) & 3.77(3) & 3.91(2) \\
 \hline
 \end{tabular}
\caption{Parameters characterizing the simulations.
\label{tab:par}}
\end{table}

\begin{table}
\begin{tabular}{c c c c c c c c}
\hline
$\kappa$ & $a\,m_q$ & $a\,M_\pi$ & $a\,F_\pi$ & $a\,m_V$   &  $a\,m_B(S=\frac{3}{2})$ & $a\,m_B(S=\frac{1}{2})$ \\
\hline
0.1220 & 0.148 & 0.598(1) &  0.077(1) &  0.676(2)  &  1.089(6) &  1.035(5) \\
0.1230 & 0.119 & 0.527(1) &  0.072(1) &  0.619(3)  &  1.007(7) &  0.942(6) \\
0.1240 & 0.089 & 0.449(2) &  0.067(1) &  0.554(4)  &  0.926(10) &  0.845(8) \\
0.1245 & 0.074 & 0.407(2) &  0.064(1) &  0.522(5)  &  0.886(11) &  0.795(8) \\
0.1250 & 0.059 & 0.361(2) &  0.062(1) &  0.490(6)  &  0.846(13) &  0.744(10) \\
0.1253 & 0.050 & 0.331(2) &  0.059(1) &  0.470(7)  &  0.821(15) &  0.711(10) \\
0.1257 & 0.038 & 0.288(2) &  0.056(1) &  0.449(8)  &  0.786(18) &  0.663(12) \\
0.1260 & 0.029 & 0.253(2) &  0.054(1) &  0.428(11)  &  0.757(19) &  0.621(13) \\
0.1261 & 0.026 & 0.240(2) &  0.054(1) &  0.421(12)  &  0.747(20) &  0.606(13) \\
0.1262 & 0.022 & 0.225(3) &  0.053(1) &  0.445(9)  &  0.736(20) &  0.589(14) \\
0.1265 & 0.013 & 0.177(3) &  0.052(1) &  0.428(12)  &  0.697(24) &  0.527(17) \\
0.1266 & 0.010 & 0.156(4) &  0.051(1) &  0.421(14)  &  0.677(29) &  0.495(22) \\
0.1267 & 0.006 & 0.133(6) &  0.051(1) &  0.432(13)  &  0.643(43) &  0.449(40) \\
\hline
 \end{tabular}
\caption{ Masses in lattice units for the $SU(3)$ data sets. From left to right, the entries
are the  hopping parameter $\kappa$, the Axial Ward Identity quark mass, the pseudoscalar mass, the
pseudoscalar decay constant, the vector meson mass, and the baryons, labeled by their spin $S$.
\label{tab:su3}}
\end{table}

\begin{table}
\begin{tabular}{c c c c c c c c}
\hline
$\kappa$ & $a\,m_q$ & $a\,M_\pi$ & $a\,F_\pi$ & $a\,m_V$ & $a\,m_B(S=\frac{5}{2})$ &  $a\,m_B(S=\frac{3}{2})$ & $a\,m_B(S=\frac{1}{2})$ \\
\hline
0.1240 & 0.127 & 0.565(1) &  0.097(1) &  0.655(1)  &  1.866(7) &  1.817(6) &  1.787(6) \\
0.1250 & 0.098 & 0.488(1) &  0.090(1) &  0.593(1)  &  1.711(7) &  1.652(6) &  1.617(6) \\
0.1260 & 0.070 & 0.403(1) &  0.083(1) &  0.532(2)  &  1.562(7) &  1.490(7) &  1.448(6) \\
0.1265 & 0.055 & 0.356(1) &  0.079(1) &  0.500(2)  &  1.482(8) &  1.402(7) &  1.355(6) \\
0.1270 & 0.041 & 0.302(2) &  0.073(1) &  0.469(3)  &  1.419(11) &  1.324(8) &  1.270(7) \\
0.1275 & 0.026 & 0.240(2) &  0.069(1) &  0.440(4)  &  1.361(10) &  1.250(13) &  1.184(10) \\
0.1278 & 0.017 & 0.193(3) &  0.065(1) &  0.424(5)  &  1.284(12) &  1.163(15) &  1.085(13) \\
0.1280 & 0.009 & 0.155(7) &  0.063(1) &  0.413(7)  &  1.247(20) &  1.105(36) &  1.006(31) \\
  \\
\hline
\end{tabular}
\caption{ Masses in lattice units for the $SU(5)$ data sets.
\label{tab:su5}}
\end{table}
\begin{table}
\begin{tabular}{c c c c c c c c c}
\hline
$\kappa$ & $a\,m_q$& $a\,M_\pi$ & $a\,F_\pi$ & $a\,m_V$ & $a\,m_B(S=\frac{7}{2})$ & $a\,m_B(S=\frac{5}{2})$ &  $a\,m_B(S=\frac{3}{2})$ & $a\,m_B(S=\frac{1}{2})$\\
\hline
0.1260 & 0.115 & 0.565(1) &  0.115(1) &  0.663(1)  &  2.668(11) &  2.625(10) &  2.595(10) &  2.577(10) \\
0.1270 & 0.088 & 0.488(1) &  0.107(1) &  0.603(1)  &  2.471(13) &  2.420(12) &  2.383(11) &  2.361(11) \\
0.1280 & 0.062 & 0.401(1) &  0.097(1) &  0.537(2)  &  2.273(18) &  2.213(13) &  2.166(13) &  2.139(12) \\
0.1290 & 0.036 & 0.299(1) &  0.086(1) &  0.471(3)  &  2.075(28) &  1.998(17) &  1.938(17) &  1.904(18) \\
0.1295 & 0.022 & 0.235(2) &  0.081(1) &  0.438(4)  &  1.967(27) &  1.891(20) &  1.818(21) &  1.777(22) \\
0.1297 & 0.017 & 0.205(2) &  0.078(1) &  0.426(5)  &  1.924(34) &  1.843(26) &  1.773(26) &   \\
0.1298 & 0.013 & 0.178(3) &  0.076(1) &  0.427(4)  &  1.890(45) &  1.833(40) &  1.735(38) &   \\
\hline
\end{tabular}
\caption{ Masses in lattice units for the $SU(7)$ data sets.
\label{tab:su7}}
\end{table}
\begin{center}
\begin{table}
\hspace{-2cm}
\begin{minipage}{1.5in}
\begin{center}
\hspace*{0cm}
\begin{tabular}{c c }
\hline
$\kappa$ & $\Delta m_B (\frac{3}{2},\frac{1}{2}) $\\
\hline
0.1220 & 0.054(3)  \\
0.1230 & 0.065(4)  \\
0.1240 & 0.080(6)  \\
0.1245 & 0.091(7)  \\
0.1250 & 0.102(8)  \\
0.1253 & 0.110(10)  \\
0.1257 & 0.123(12)  \\
0.1260 & 0.136(14)  \\
0.1261 & 0.141(14)  \\
0.1262 & 0.147(15)  \\
0.1265 & 0.170(21)  \\
0.1266 & 0.181(28)  \\
0.1267 & 0.195(49)  \\
\hline
 \end{tabular}
\caption{ HF splittings  for $N_c=3$
\label{tab:su3HF}}
\end{center}
\end{minipage}
\hspace*{2cm}
\begin{minipage}{3in}
 \begin{tabular}{c c c c}
\hline
$\kappa$ &  $\Delta m_B (\frac{5}{2},\frac{3}{2}) $ & $\Delta m_B (\frac{5}{2},\frac{1}{2}) $ & $\Delta m_B (\frac{3}{2},\frac{1}{2}) $\\
\hline
0.1240 & 0.050(2)  & 0.080(2 )& 0.030(1) \\
0.1250 & 0.059(2)  & 0.094(3 )& 0.035(1) \\
0.1260 & 0.071(2)  & 0.114(4 )& 0.043(2) \\
0.1265 & 0.080(3)  & 0.127(5 )& 0.047(2) \\
0.1270 & 0.096(7)  & 0.150(9 )& 0.054(4) \\
0.1275 & 0.112(11)  & 0.177(10 )& 0.066(12) \\
0.1278 & 0.121(13)  & 0.200(13 )& 0.079(11) \\
0.1280 & 0.142(34)  & 0.241(37 )& 0.099(45) \\
\hline
 \end{tabular}
\caption{ HF splittings for $N_c=5$
\label{tab:su5HF}}
\vspace{.5cm}
\hspace*{0cm}
\begin{tabular}{c c c c c  }
\hline
$\kappa$ & $\Delta m_B (\frac{7}{2},\frac{5}{2}) $&$\Delta m_B (\frac{7}{2},\frac{3}{2}) $&$\Delta m_B (\frac{5}{2},\frac{3}{2}) $&$\Delta m_B (\frac{5}{2},\frac{1}{2}) $ \\
\hline
0.1260 & 0.042(4)  & 0.072(6) & 0.030(2)  & 0.019(1) \\
0.1270 & 0.051(4)  & 0.088(7) & 0.037(3)  & 0.022(1) \\
0.1280 & 0.060(13)  & 0.106(13) & 0.046(5)  & 0.027(2) \\
0.1290 & 0.077(22)  & 0.136(24) & 0.059(8)  & 0.035(4) \\
0.1295 & 0.076(21)  & 0.149(20) & 0.073(14)  & 0.041(14) \\
0.1297 & 0.081(22)  & 0.151(27) & 0.071(14)  &  \\
0.1298 & 0.057(42)  & 0.156(45) & 0.098(36)  &  \\
\hline
 \end{tabular}
\caption{ HF splittings for $N_c=7$.
\label{tab:su7HF}}
\end{minipage}
\end{table}
\end{center}

The data are extended from that of  Ref.~\cite{DeGrand:2012hd} in two ways.
First,   the $N_c=7$, $S=\frac{1}{2}$ baryon was added to the set of measured states. As described
 in Ref.~\cite{DeGrand:2012hd}, lower $S$ states contain many more contractions of creation 
and annihilation operators
into propagators. The new baryon's propagator has about 1.5 million determinants needed to be evaluated, per site.
 Second,  the spectroscopy has been extended
for all $N_c$'s to lower quark mass. Comparisons of large-$N_c$ predictions do not necessary require small quark mass,
but of course chiral extrapolations need the lightest possible quark masses. The quark mass was lowered   until
at very small quark mass   ``exceptional configurations'' are encountered.  These are gauge configurations
on which the Dirac operator has eigenvalues close to the value zero, so it becomes ill-conditioned, and hence
non-invertible. This situation is well-known from quenched studies in $N_c=3$. There, it is cured by doing simulations
with dynamical fermions, since the zero modes mean that the fermion determinant is zero and these configurations
 never appear in the Markov chain.

In any lattice calculation, the values of two observables computed on the same set
 of lattice configurations are highly correlated. In the case of mass differences, this means that
the uncertainty in  $m(S)-m(S')$ is smaller than what a naive combination of the individual uncertainties
in $m(S)$ and $m(S')$ would indicate.  The baryon masses and the mass differences were simultaneously computed 
by performing a single-elimination jackknife analysis of the appropriate correlators.
This ensures that the difference of masses  (as recorded, for example, in Table  \ref{tab:su3}) is equal to the mass differences
(as recorded in Table \ref{tab:su3HF}).

\section{The Effective Theory Analysis of LQCD  Results}    \label{sec:Analysis}

This section presents the analysis of the LQCD results of the previous section in the framework of the combined HBChPT and $1/N_c$ expansions with the $\xi$ power counting~\cite{CalleCordon:2012xz}. This analysis should be considered as a first attempt at determining the $N_c$ dependencies of baryon masses in that expansion  from  the  first  LQCD results for baryons at varying $N_c$. The main limitation of the present analysis is the current range of quark masses in the simulations, which corresponds to  350~MeV~$\leq M_\pi\leq$~1400~MeV. Due to that limitation and in order to have more information at lower values of $M_\pi$,  inputs from other LQCD calculations at $N_c=3$ will be included~\cite{Aoki:2008sm,WalkerLoud:2008bp,Alexandrou:2011py}. This, however, introduces a different  limitation, which is the matching  between full QCD results at $N_c=3$ and the results of the previous section performed in the quenched approximation. For  quark masses in the domain of Ref~\cite{DeGrand:2012hd},  quenching effects are expected to be relatively small. It is obvious that for a more accurate analysis the simulations at varying $N_c$ must be carried out at lower quark masses, where at the same time dynamical quarks will be needed, which is perhaps at this time a very challenging task. Nevertheless, at the present level of accuracies such a combination seems to be justified as it is shown by the following  discussion.

In general, the effects of quenching on hadron observables are suppressed by a factor $1/N_c$, as it follows from 't Hooft's $N_c$ power counting (exceptions exist, such as   the topological susceptibility which is $\ord{N_c^0}$ in the quenched approximation and vanishes in  the chiral limit in full QCD \cite{Witten:1979vv}). In baryons, quenching affects baryon masses at $\ord{N_c^0}$, but this effect respects spin-flavor symmetry, while in the HF splittings it is an effect $\ord{1/N_c^2}$. Thus, to be rigorously consistent with sub-leading orders in the $1/N_c$ expansion, the LQCD calculations will need to be in full QCD, which is of course increasingly expensive with $N_c$. Nonetheless, QCD in the quenched approximation itself  admits a $1/N_c$ expansion, and it is therefore of interest in its own right.

The issue of combining quenched and unquenched data is now analyzed. The  combination of   $N_c=3$  quenched and unquenched results  is  found to work remarkably well  when quantities are considered in units of the corresponding  $\mathring{F}_0$, defined as the chiral limit value of $F_0\equiv F_\pi\sqrt{3/N_c}$.  In particular, for the quantities of relevance to the present analysis, namely
$F_0/\mathring{F}_0~{\rm vs}~M_\pi/\mathring{F}_0$  and  $M_B/\mathring{F}_0~{\rm vs}~M_\pi/\mathring{F}_0$, there is  agreement  between the quenched and the unquenched results within the current LQCD errors, as is shown below.
\begin{center}
\begin{figure}[h!!!!]
\includegraphics[width=8.cm,angle=0]{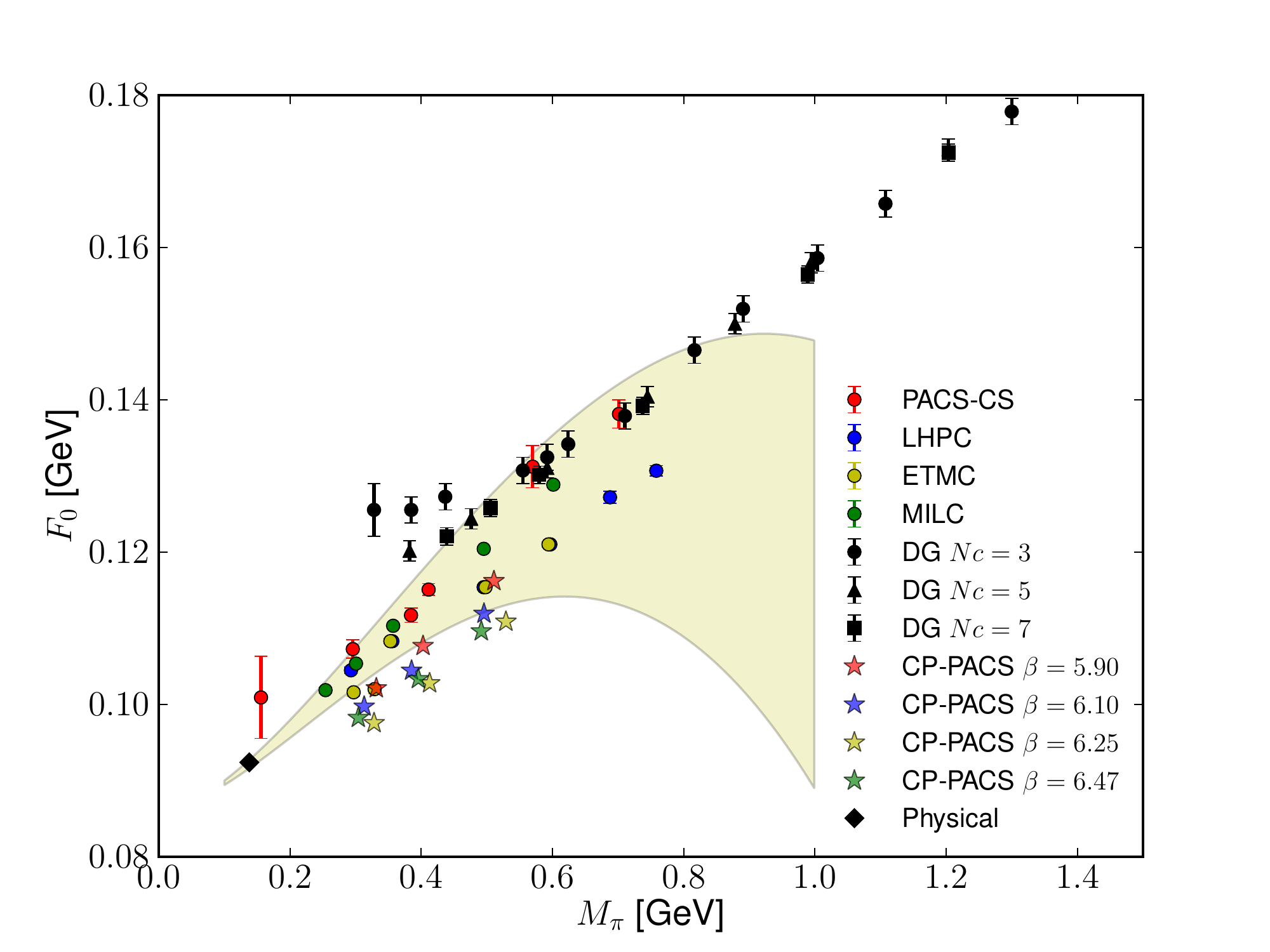}
\includegraphics[width=8.cm,angle=0]{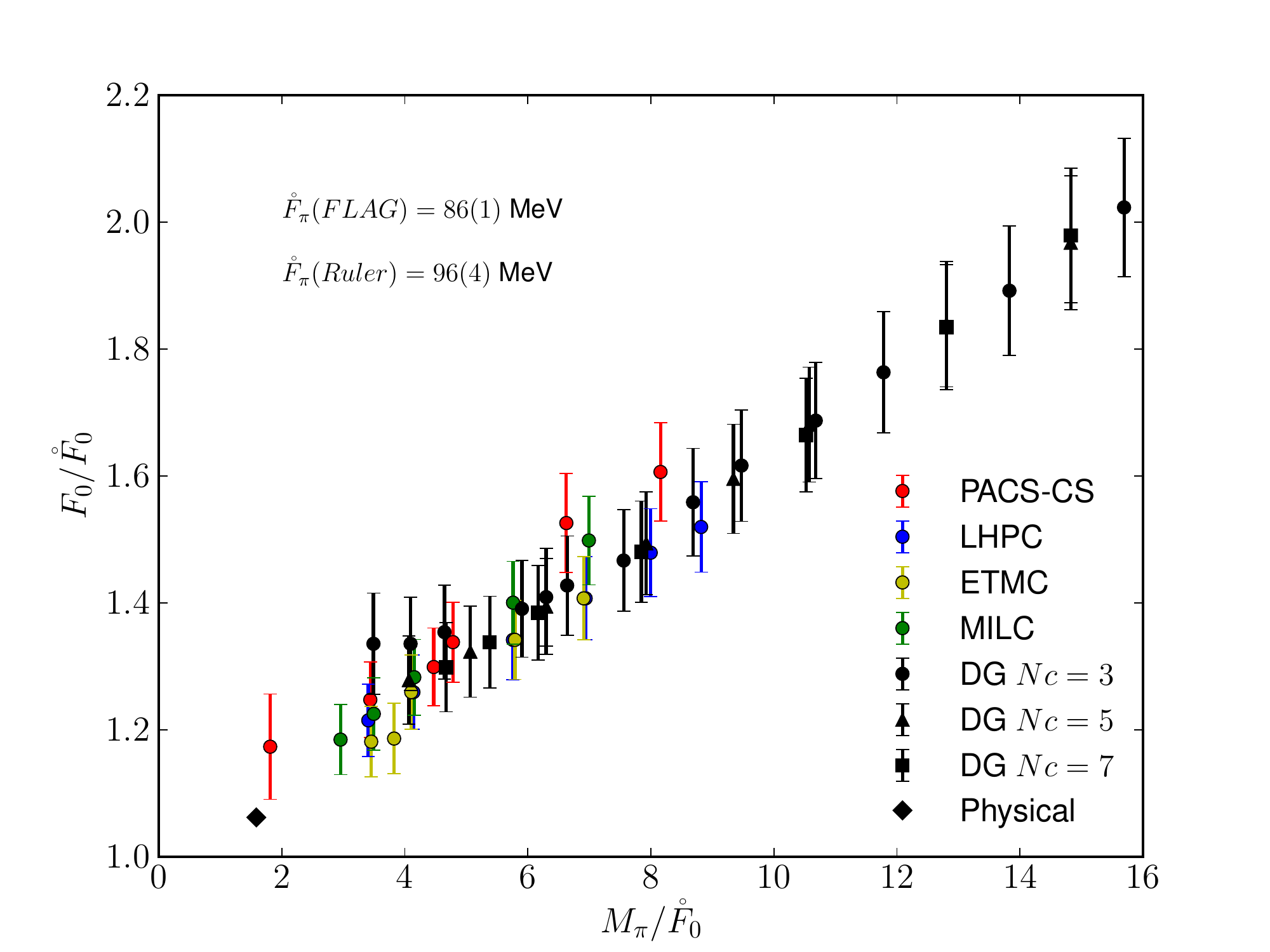}
\caption{
Left panel: $F_0\equiv F_\pi \sqrt{3/N_c}$ vs $M_\pi$ in physical units for the quenched results from CP-PACS~\cite{Aoki:2002fd} and $N_c=3,5,7$ of Ref.~\cite{DeGrand:2012hd}, and the unquenched results from PACS-CS~\cite{Aoki:2008sm}, LHPC~\cite{WalkerLoud:2008bp}, ETMC~\cite{Alexandrou:2011py} and MILC~\cite{Aubin:2004fs} collaborations. The diamond represents the physical point.
The yellow band represents $\ord{p^4}$ ChPT~\cite{Gasser:1983yg} with the recent FLAG's determination of the LECs $\bar\ell_4$ and $\mathring F_\pi$ from Eqs.~\eqref{eq:l4-FLAG} and~\eqref{eq:Fpi0-FLAG} (see Ref.~\cite{Aoki:2013ldr}). 
Right panel: $F_0/\mathring F_0$ vs $M_\pi/\mathring F_0$.  $\mathring F_0$ is determined from the FLAG's averaged value in the case of unquenched LQCD results and   the linear extrapolation value in the case of the $N_c=3,5,7$ results of Ref.~\cite{DeGrand:2012hd}.
}
\label{fig:FpivsMpi}
\end{figure}
\end{center}

First the case of pions is discussed at the level of the two quantities of relevance for this work, namely $M_\pi$ and $F_\pi$.
Since $\mathring F_0$ is crucial in the present analysis,  a detailed discussion is given of its
 determination and consistency for the different LQCD calculations being considered.
 In Fig.~\ref{fig:FpivsMpi}, $F_0$ vs. $M_\pi$ is displayed for several LQCD calculations.
 The left panel displays results from different LQCD collaborations in quenched~(DeGrand and CP-PACS)
 and full QCD~(PACS-CS, LHPC, ETMC and MILC). Results are displayed using the quoted lattice spacings 
provided by the different collaborations, namely,
$a\simeq 0.08~{\rm fm}$ (DeGrand)~\cite{DeGrand:2012hd},
$a\simeq 0.09~{\rm fm}$ (PACS-CS)~\cite{Aoki:2008sm},
$a\simeq 0.124~{\rm fm}$ (LHPC)~\cite{WalkerLoud:2008bp},
$a\simeq 0.08-0.124~{\rm fm}$ (ETMC)~\cite{Alexandrou:2011py},
$a\simeq 0.124~{\rm fm}$ (MILC)~\cite{Aubin:2004fs}, and
$a\simeq 0.05-0.1~{\rm fm}$ (CP-PACS)~\cite{Aoki:2002fd}, 
where the intervals indicate the range of lattice spacings where the  simulations were carried out.

At low quark mass, the quenched data sets show considerable differences with each other and with the unquenched sets.
On the other hand,  it is expected that  the quenched approximation should approach the unquenched
 results at larger quark mass where quark loops  become dynamically suppressed.  This characteristic is
 clearly met by the  PACS-CS, MILC and DeGrand results.
Moreover, according to the  $1/N_c$ expansion,  at large $N_c$ the plot $F_0$ vs $M_\pi$ must lie
on a universal curve. This is clearly  illustrated in Fig.~\ref{fig:FpivsMpi}, 
where the results for $N_c=3$, $5$ and $7$ fall on an approximately common curve.
 This has also been shown in other LQCD calculations at varying $N_c$~\cite{Lucini:2012gg,Bali:2013kia}.

In order to properly carry out a comparison between results from different collaborations, in the
 analysis one could use the physical energy scale determined from the lattice and transform everything to
 physical units.
However, this may introduce  dependencies on lattice artifacts,  which seem to be 
 evident from Fig.~\ref{fig:FpivsMpi} for the different unquenched  calculations. 
In order to avoid this issue of scale matching between calculations, it  is convenient to
 compare  dimensionless ratios of observables. 
As mentioned earlier, $ \mathring F_0 (N_c)$ will be  used  to define these dimensionless ratios.
Recently, the FLAG collaboration \cite{Aoki:2013ldr} has carried out a detailed study 
 providing average values, namely:
\bea
\label{eq:Fpi0-FLAG}
\mathring F_\pi\ \rm{(FLAG)}  & = & 86 \pm 1 \ \rm{MeV} \, , \\
\label{eq:l4-FLAG}
\bar{l}_4\ \rm{(FLAG)} &=& 4.4 \pm 0.4 \ .
\eea
In the present work only  a subset of lattice data sets is used.  In order to check consistency with the   world averages, an analysis of those sets is performed.
In the unquenched case, for $N_c=3$, $ \mathring F_0 $ is obtained from lattice results 
 using the second order results in ChPT~\cite{Gasser:1983yg},
 \bea
M_\pi^2 &=& \mathring M_\pi^2 \left(1 + \frac{\mathring M_\pi^2}{32 \pi^2 \mathring F_\pi} 
\big( \log\frac{\mathring M_\pi^2}{\mu ^2}+ 64 \pi^2 l_3(\mu) \big)  \right)\, ,\\
\label{eq:GL-ChPT-Mpi}
F_\pi &=& \mathring F_\pi \left( 1 - \frac{\mathring M_\pi^2}{16 \pi^2 \mathring F_\pi}
 \big( \log{\frac{\mathring M_\pi^2}{\mu ^2}} - 64 \pi^2 l_4(\mu) \big)  \right)\, ,
\label{eq:GL-ChPT-Fpi}
\eea
where $\mu$ is the renormalization scale and $\mathring M_\pi$ and $\mathring F_\pi$ are the lowest order
 values of these quantities; in particular $\mathring M_\pi^2 = 2 B m_q$. The LECs are given by
\bea
l_3(\mu) &=& - \frac{1}{64 \pi^2} \big( \bar{l_3} + \log{( (138 \rm MeV)^2/\mu[\rm MeV] ^2 ) } \big)\, ,\\
l_4(\mu) &=& \frac{1}{64 \pi^2} \big( \bar{l_4} + \log{( (138 \rm MeV)^2/\mu[\rm MeV] ^2 ) } \big)\, ,
\eea
where from phenomenology 
and   recent LQCD calculations 
 the values of the LECs $\bar l_3 = 3.16$ and  $\bar l_4 = 4.03$ are obtained.

To the order of accuracy needed,  $\mathring M_\pi$ can be  replaced by  $M_\pi$ in Eq.~\eqref{eq:GL-ChPT-Fpi}. In addition, the ratio $\mathring M_\pi/M_\pi$ remains close to unity for the whole range of pion masses. The results of the fits to unquenched LQCD are shown in Table~\ref{tab:results-Fpi-unquenched}.

\begin{table}[htdp]
\caption{Chiral limit extrapolated $\mathring F_\pi$  from the unquenched LQCD results used in the fits. 
The quoted lattice spacings from the different collaborations are used to convert to physical units. 
The $\ord{p^4}$ formula for $F_\pi$, Eq.~\eqref{eq:GL-ChPT-Fpi}, is used to 
obtain $\mathring F_\pi$ and $\bar{l_4}$,   replacing  $\mathring M_\pi$ by $M_\pi$ as given by  the lattice results.
}
\begin{center}
\begin{tabular}{c|c|c|c}\hline\hline
LQCD Collaboration & $\chi^2_{DOF}$ & $\mathring F_\pi$ [MeV] & $\bar{l_4}$ \\\hline
PACS-CS~\cite{Aoki:2008sm}    & 1.49 & $ 88.3 \pm 1.2 $& $4.4  \pm 0.1$  \\
LHPC~\cite{WalkerLoud:2008bp} & 0.40 &$ 88.4 \pm 0.5$ &$ 4.09  \pm 0.03$ \\
ETMC~\cite{Alexandrou:2011py} & 1.46 & $ 88.9 \pm 1.2$ &$ 3.50  \pm 0.16$\\
MILC~\cite{Aubin:2004fs}      & 0.48 & $86.5 \pm 0.3$ & $4.45  \pm 0.02$ \\
\hline\hline
\end{tabular}
\end{center}
\label{tab:results-Fpi-unquenched}
\end{table}

The results in Table~\ref{tab:results-Fpi-unquenched} are quite compatible with the FLAG ones. With that agreement it is reasonable to
 choose a common $\mathring F_\pi$  given in  Eq.~\eqref{eq:Fpi0-FLAG} for all the unquenched LQCD results
 in what follows.

Now we turn to the quenched data set. Recall that we need a value for  $ \mathring F_0 $ to scale our baryon masses.  The difference between the quenched data sets, plus the curvature of the $N_c=3$ data of Ref.~\cite{DeGrand:2012hd} make us suspicious of the quality of the quenched data for $F_\pi$ at small quark mass. We adopt a more phenomenological approach to obtain $\mathring F_0$:
It is obvious that quenching effects become more important at small pion masses. However, the differences between quenched and unquenched results should become smaller as the pion mass increases. This effect is noticed in comparing the quenched with the PACS-CS and MILC results in the left panel of Fig.~\ref{fig:FpivsMpi}. It is then reasonable to give preference to the large pion mass results of~\cite{DeGrand:2012hd}  at varying $N_c$, where quenched and full LQCD are more alike, to guide the extraction of  $\mathring F_\pi$ in this case. The striking linear behavior suggest to use a linear function in $M_\pi$ and disregard the lowest pion masses, obtaining,
\bea
\label{eq:Fpi0-Ruler}
\mathring F_0\ \rm{(Ruler)} = 94(5) \ \rm{MeV} \, .
\eea
The error stems from  the  differences observed in the extrapolated value when changing $N_c$ from 3 to 7. This value is not very different from the unquenched value in  Eq.~\eqref{eq:Fpi0-FLAG}, and in fact gives a remarkably good agreement in the ratios of baryon masses and HF splittings obtained by the  different collaborations.

The right panel of Fig.~\ref{fig:FpivsMpi}  depicts  all LQCD results for $F_0/\mathring{F}_0$ vs $M_\pi/\mathring{ F}_0$, where for the unquenched results  $\mathring{ F}_0$ is taken as the  FLAG value Eq.~\eqref{eq:Fpi0-FLAG}, and the ruler  value  Eq.~\eqref{eq:Fpi0-Ruler} for the quenched results.  Quenching effects and/or lattice artifacts for the ratios remain small for the whole range of pion masses.

\begin{center}
\begin{figure}[h!!!]
\includegraphics[width=8.cm,angle=0]{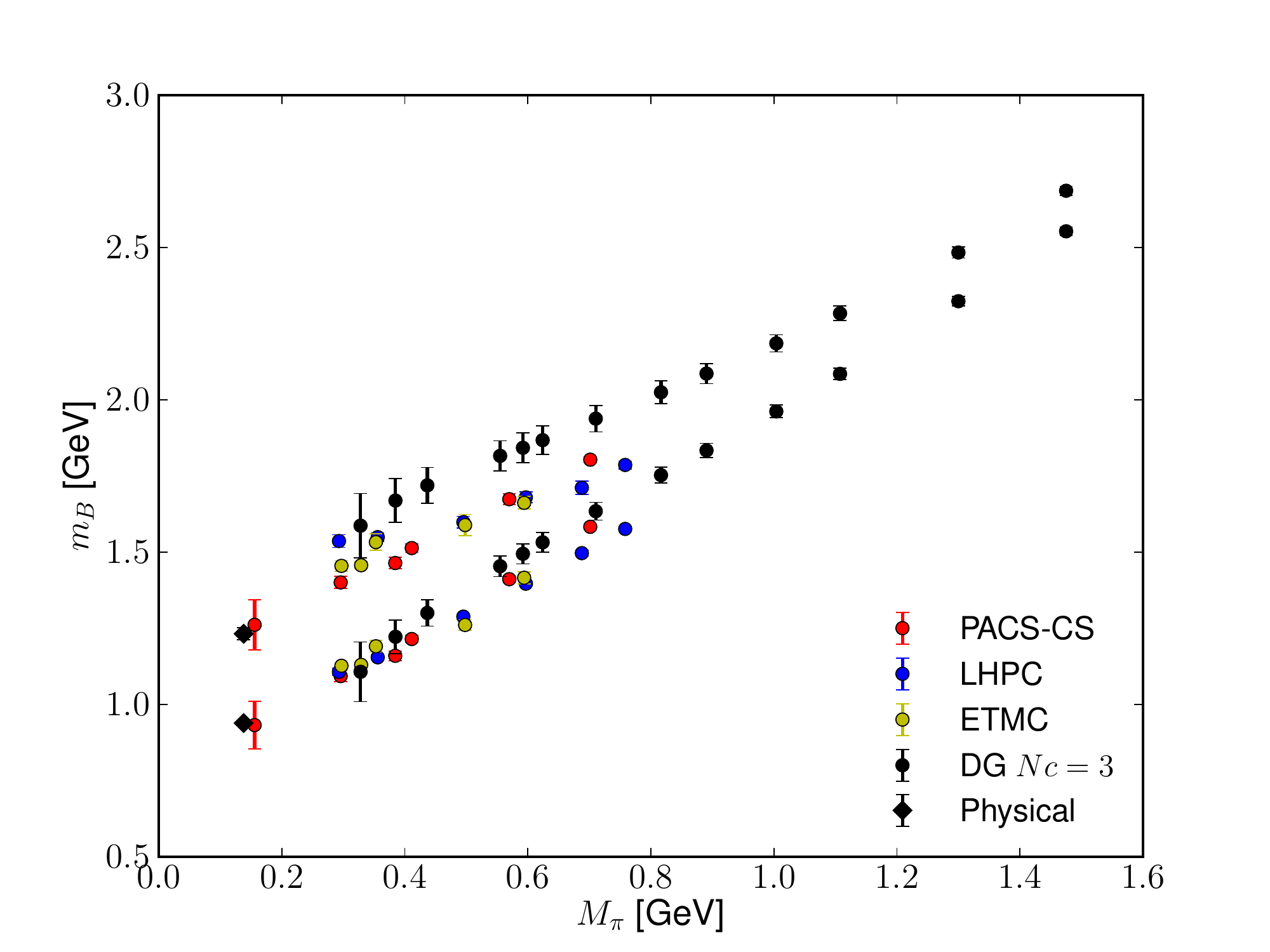}
\includegraphics[width=8.cm,angle=0]{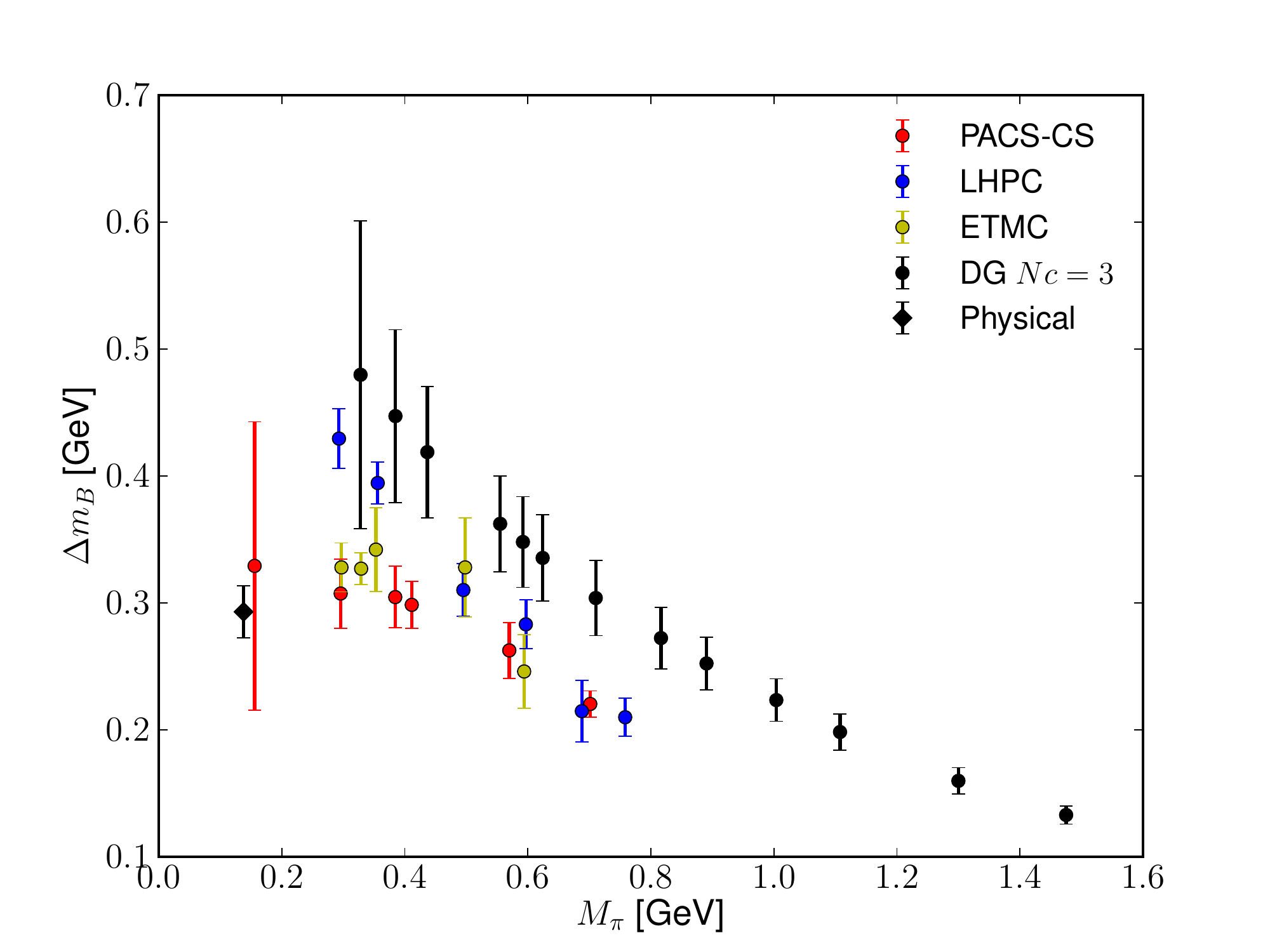}\\
\includegraphics[width=8.cm,angle=0]{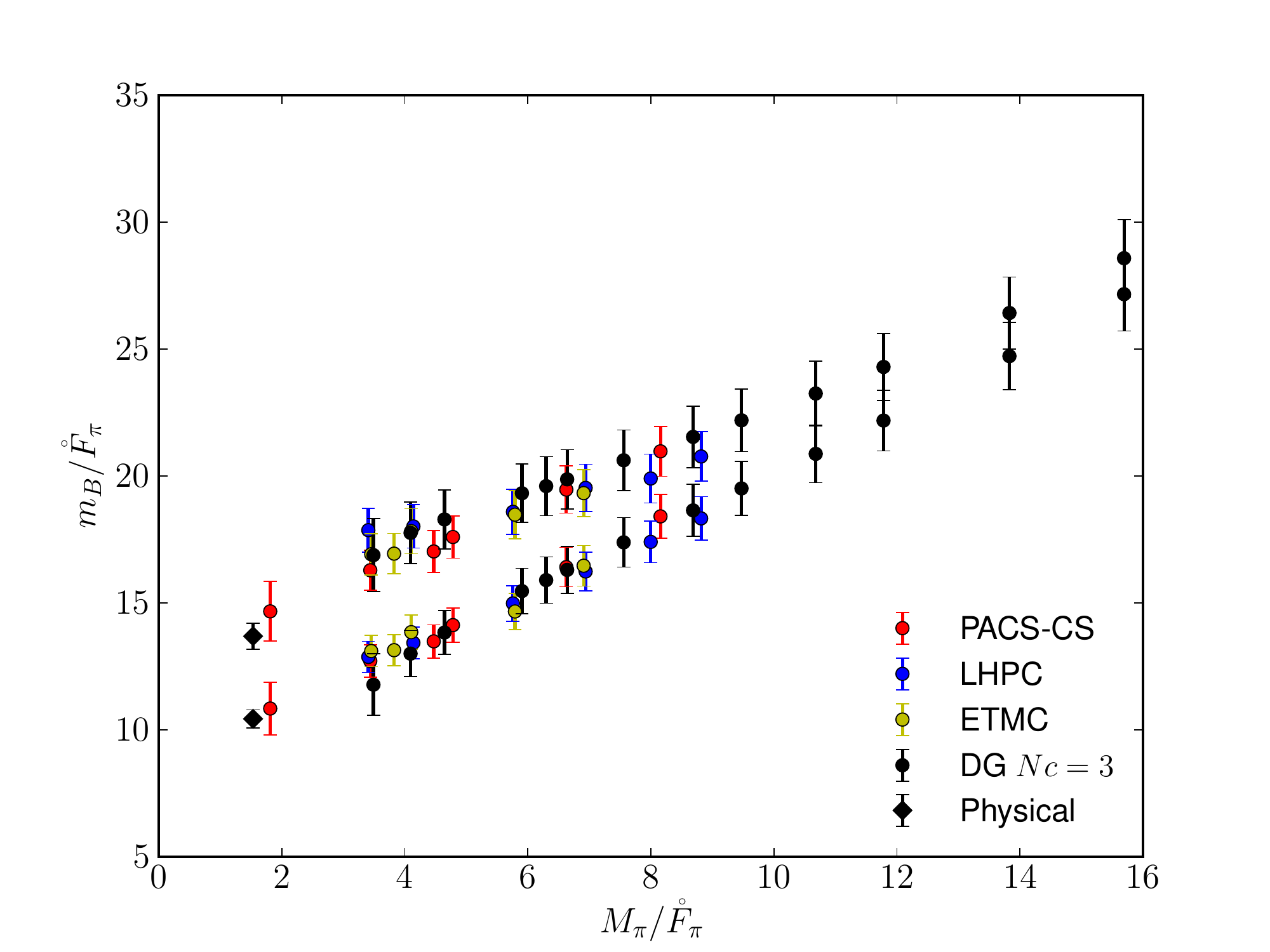}
\includegraphics[width=8.cm,angle=0]{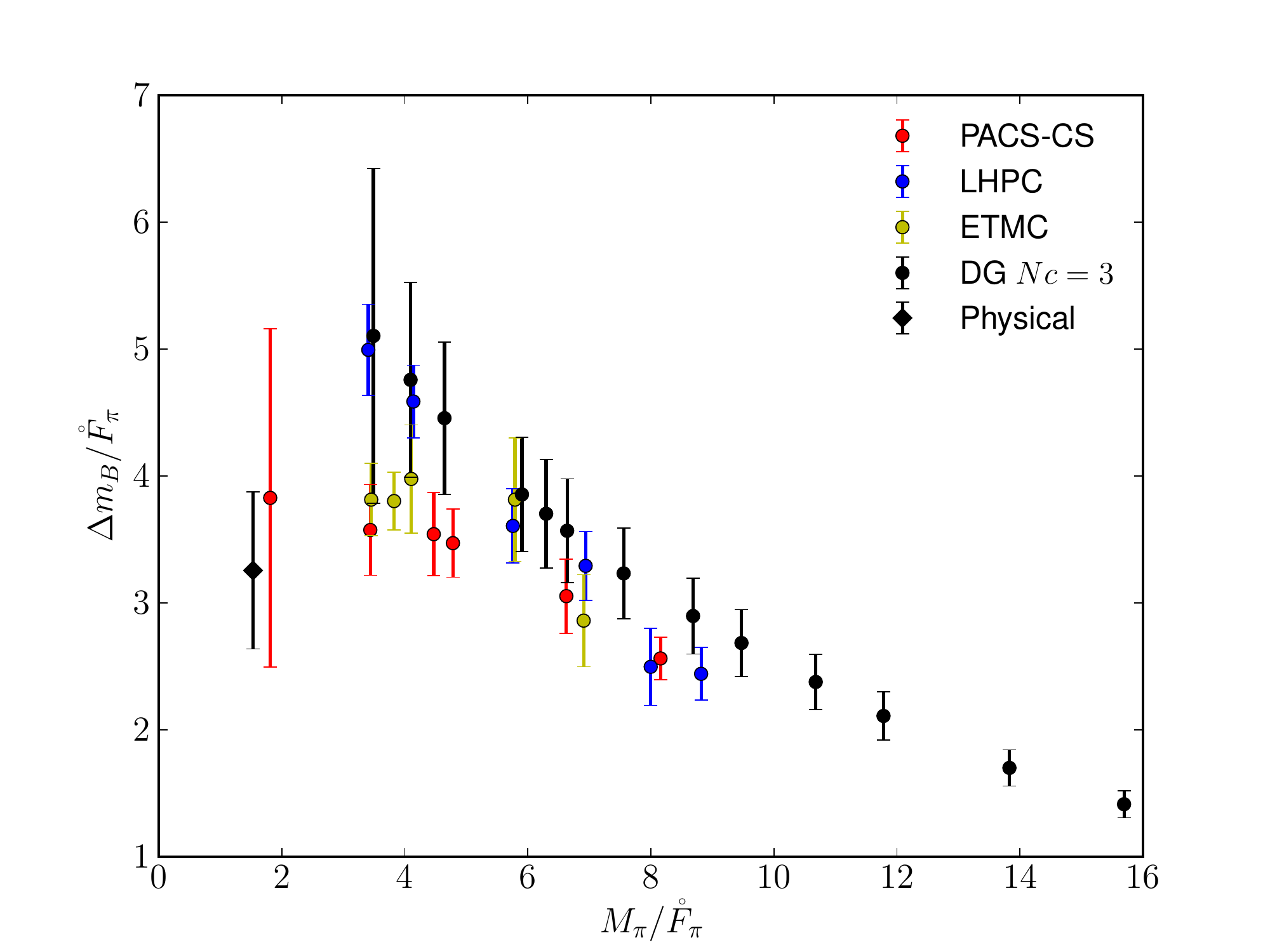}
\caption{Comparison for $N_c=3$ masses and HF splitting as functions of the pion mass of the quenched~\cite{DeGrand:2012hd} and unquenched results from different collaborations in physical units and units of $\mathring F_\pi$. Upper panels: $m_{N,\Delta}$ (left) and $m_\Delta-m_N$ (right) vs $M_\pi$ in GeV. Lower panels: $m_{N,\Delta}/\mathring F_\pi$ (left) and $(m_\Delta-m_N)/\mathring F_\pi$ (right) vs $ M_\pi/ \mathring F_\pi$.}
\label{fig:Mass-quench-effects}
\end{figure}
\end{center}

The quenching effects also become unnoticeably  small for the baryon mass ratios $m_B/\mathring F_0$ and $\Delta m_B/ \mathring F_0$ (where $\Delta m_B$ indicates a HF splitting) as shown in Fig.~\ref{fig:Mass-quench-effects}.
Notice that the HF splittings for $N_c=3$ obtained by the different LQCD collaborations give different results, although they are consistent  thanks  to the generous errors.  Thus,  because of the  current dispersion  in LQCD results with dynamical fermions,  at present there are no significant differences between quenched and unquenched results for the discussed ratios relevant to  the study carried out here.
 
In summary, although the effects of quenching can be significant for individual quantities, the ratios shown in Fig.~\ref{fig:Mass-quench-effects} involving the baryon masses are very close for the quenched and full QCD cases, justifying the approach of combining quenched and unquenched results.
 
 The next step is to carry out the  EFT analysis of the baryon mass ratio  $m_{B}/\mathring F_0$ vs $ M_\pi/ \mathring F_0$ supplemented by the  HF splittings $\Delta m_{B}(S,S')/\mathring F_0$ vs $ M_\pi/ \mathring F_0$ .
The   results are presented in Table~\ref{tab:results}, and can be briefly summarized as follows:
\begin{itemize}
 \item Fit I: is a fit to $\ord{\xi^3}$ in the expansion, where only the HF splittings from PACS-CS~\cite{Aoki:2008sm} and DeGrand's quenched results in Sec.~\ref{sec:LQCD} are included. 
 \item Fit II: is a combined fit where masses and HF splittings are both included. Results at  different orders in the $\xi$-expansion are shown.
 \item Fit III: is a $\ord{\xi^3}$ combined fit to masses and HF splittings but setting $\mathring g_A=0$. The purpose of this fit is to give a measure of the one-loop contributions in Fits I and II.
 \end{itemize}
The determination of the LECs in the  fits is made as follows: initially a fit of the baryon HF splittings is performed,  where only the spin-dependent terms in the mass formula are needed.  This is followed by a combined fit  including the masses, which allows  for the determination of the rest of the LECs. It should be emphasized that with the present inputs not all LECs can be determined, as discussed below.

One issue in the analysis is the value to be used for the axial coupling $\mathring{g}_A$, which determines the size of the one-loop contributions. Because of the lack of results for the axial coupling at varying $N_c$, it has been determined by a previous analysis from LQCD unquenched calculations at $N_c=3$  in Ref.~\cite{CalleCordon:2012xz,Cordon:2013era}, where a combined fit to LQCD masses and the axial coupling was performed. The value obtained $\mathring g_A = 1.4$ will be the one used here for all $N_c$. This corresponds to neglecting corrections $\ord{1/N_c}$ to $\mathring g_A$. Quenching effects also produce effects of that size on $\mathring g_A$. All this amounts to neglecting some $\ord{\xi^3}$ effects.  Determining these  (expected to be small)  effects will require  LQCD  calculations of $g_A$ at different values of $N_c$.

\begin{center}
\begin{figure}[h!!!]
\includegraphics[width=5. cm, height=5. cm,angle=0]{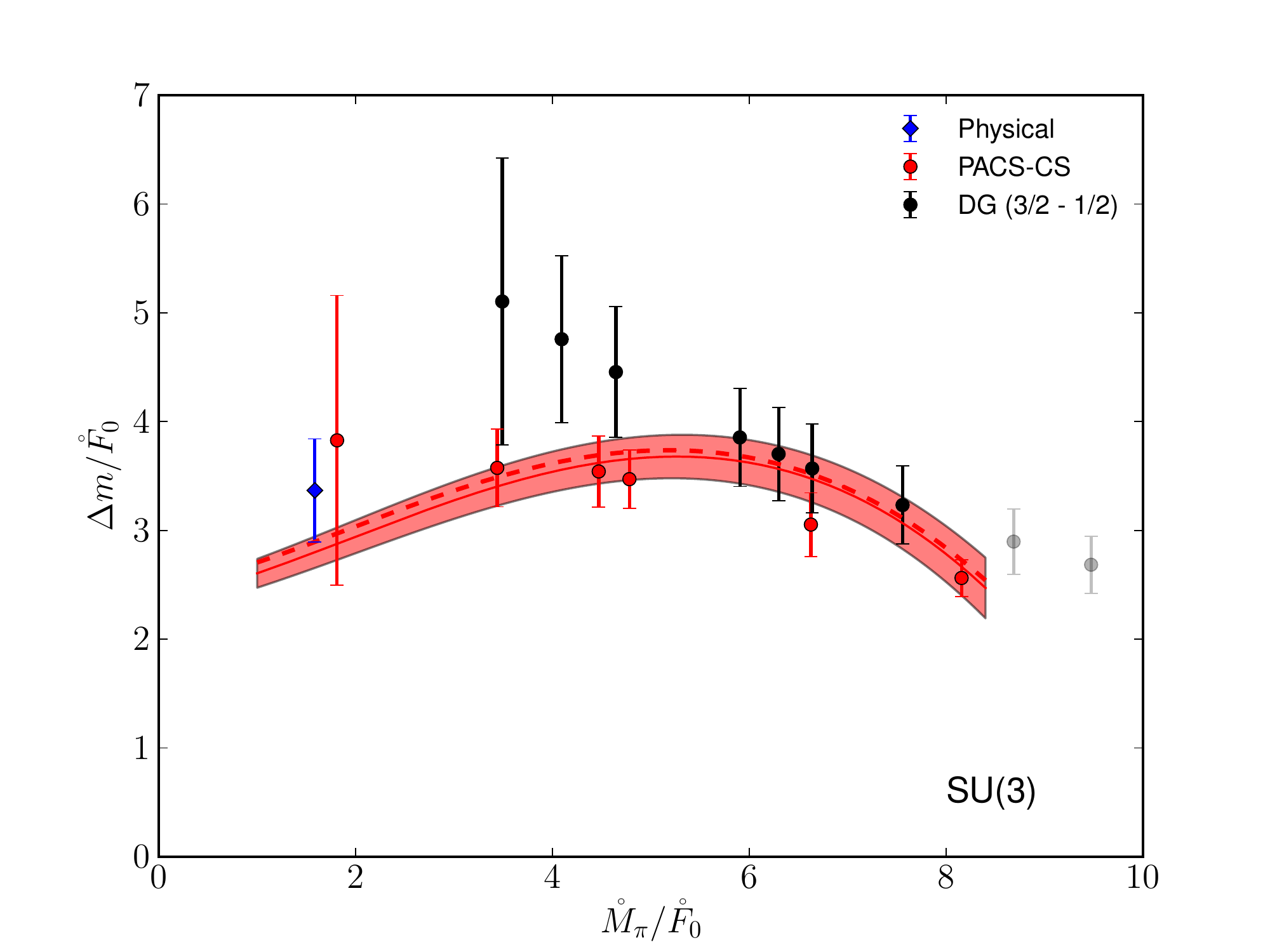}
\includegraphics[width=5. cm, height=5. cm,angle=0]{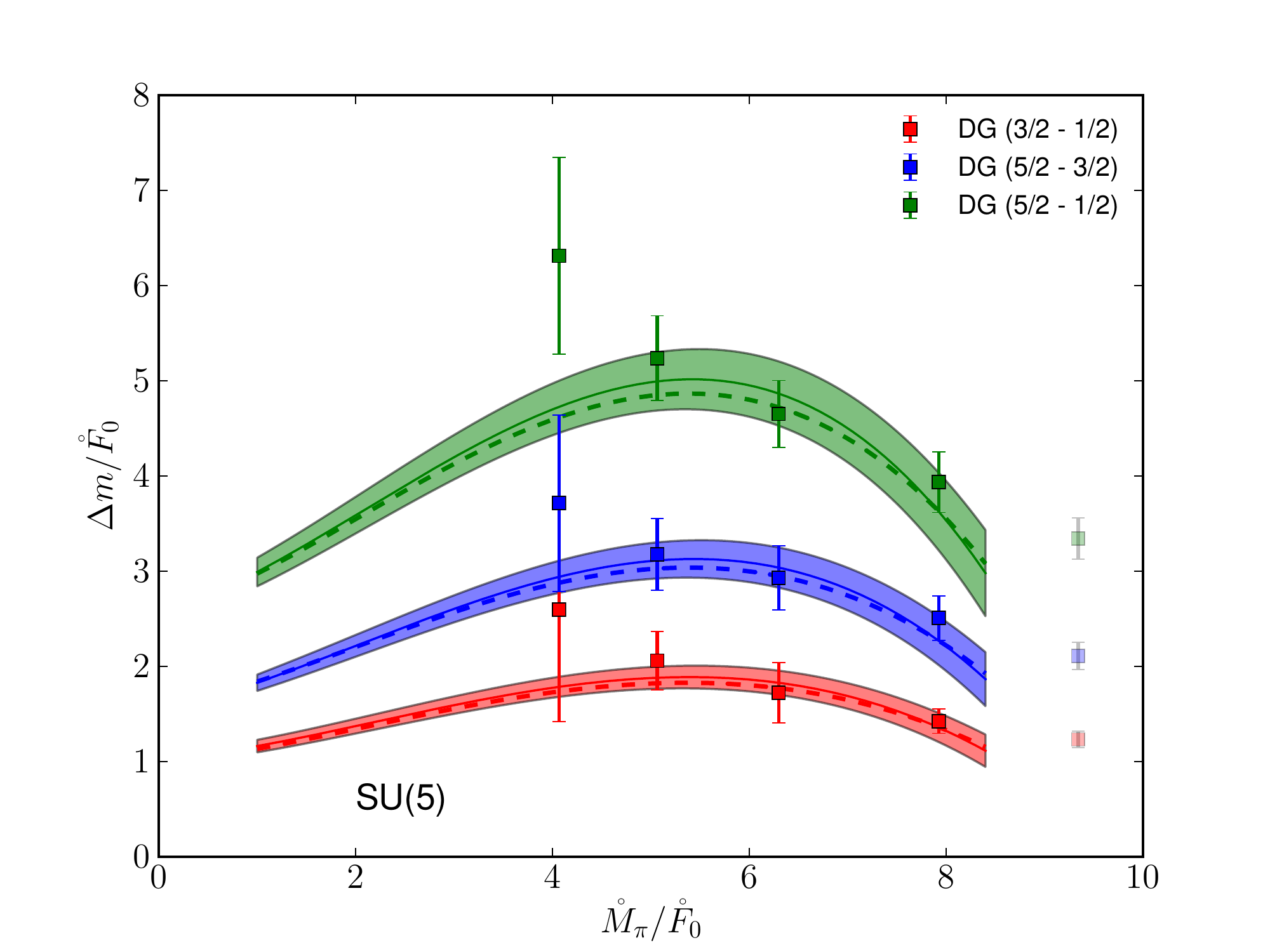}
\includegraphics[width=5. cm, height=5. cm,angle=0]{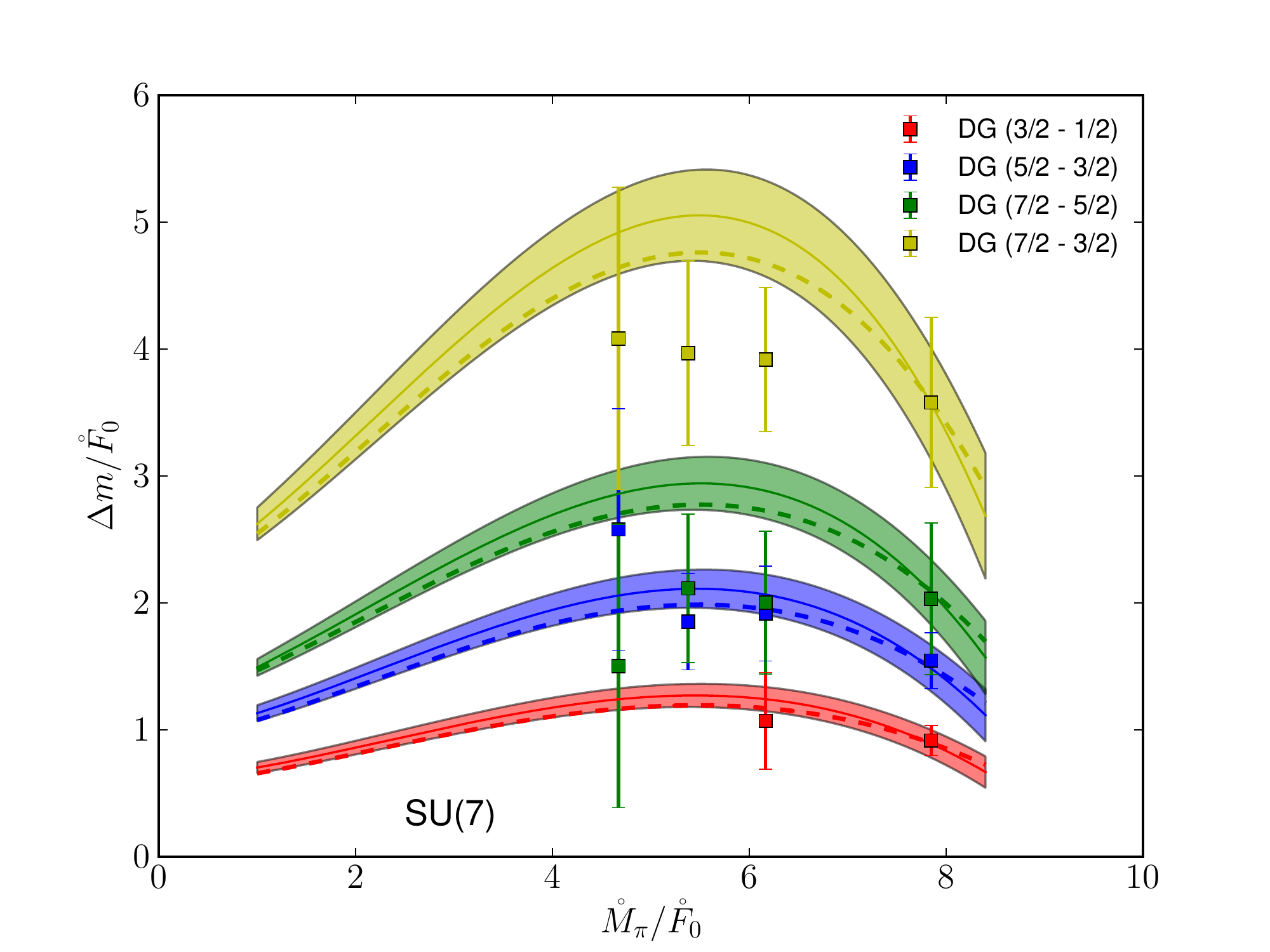}
\caption{HF splittings in the $N_c=3$, 5 and 7 multiplets. Results are in units of $\mathring{F}_0$. $N_c=3$ includes the results from PACS-CS~\cite{Aoki:2008sm}, and the  shaded points represent lattice points excluded in the fit, which correspond to  pion masses  $M_\pi \gtrsim 800$ MeV. Dashed lines for  fits when only HF lattice results are included, and the solid lines for  fits when masses are also included. }
\label{fig:MassSplitting}
\vspace*{1.cm}
\includegraphics[width=5.cm, height=5cm,angle=0]{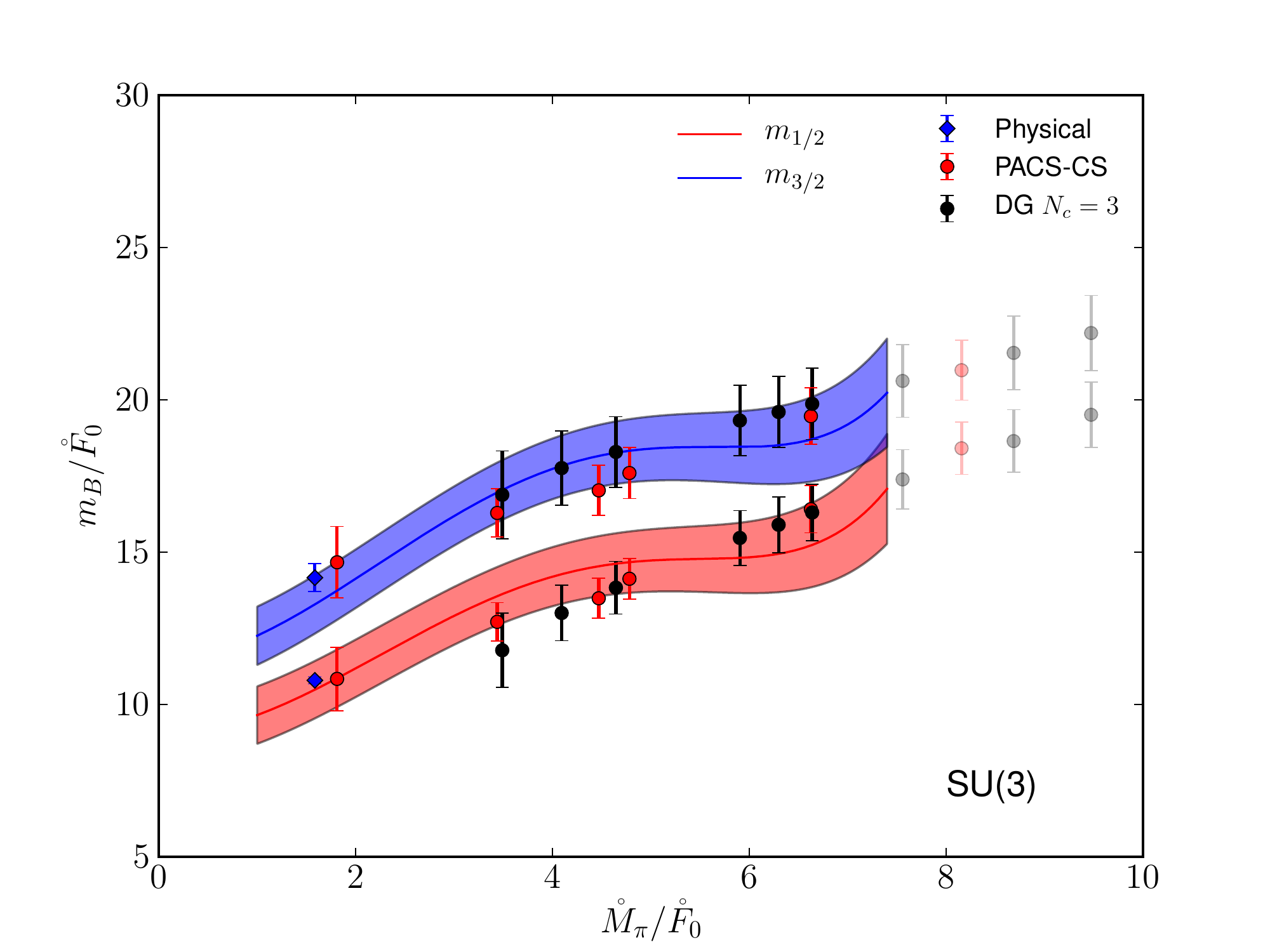}
\includegraphics[width=5.cm, height=5cm,angle=0]{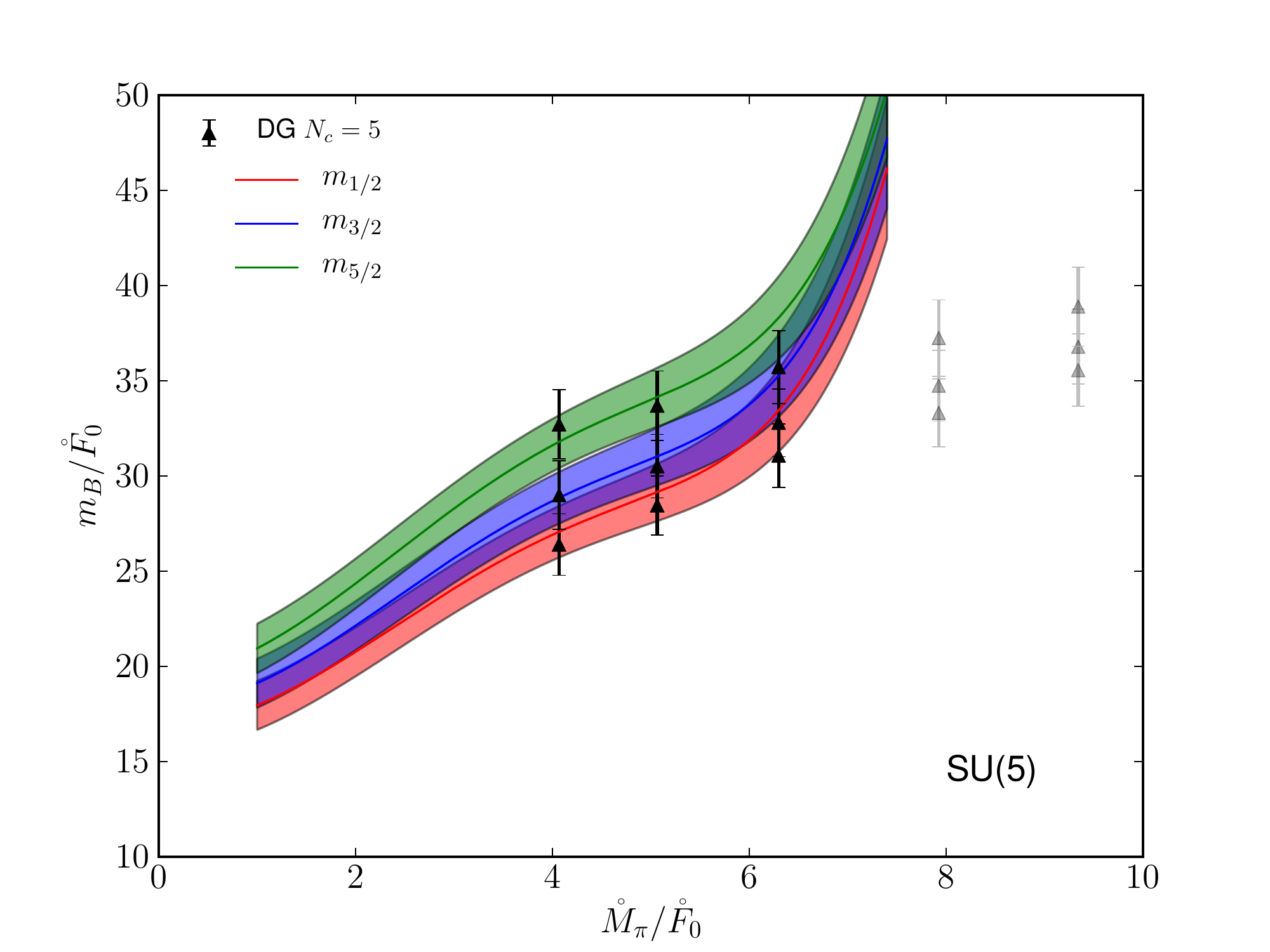}
\includegraphics[width=5.cm, height=5cm,angle=0]{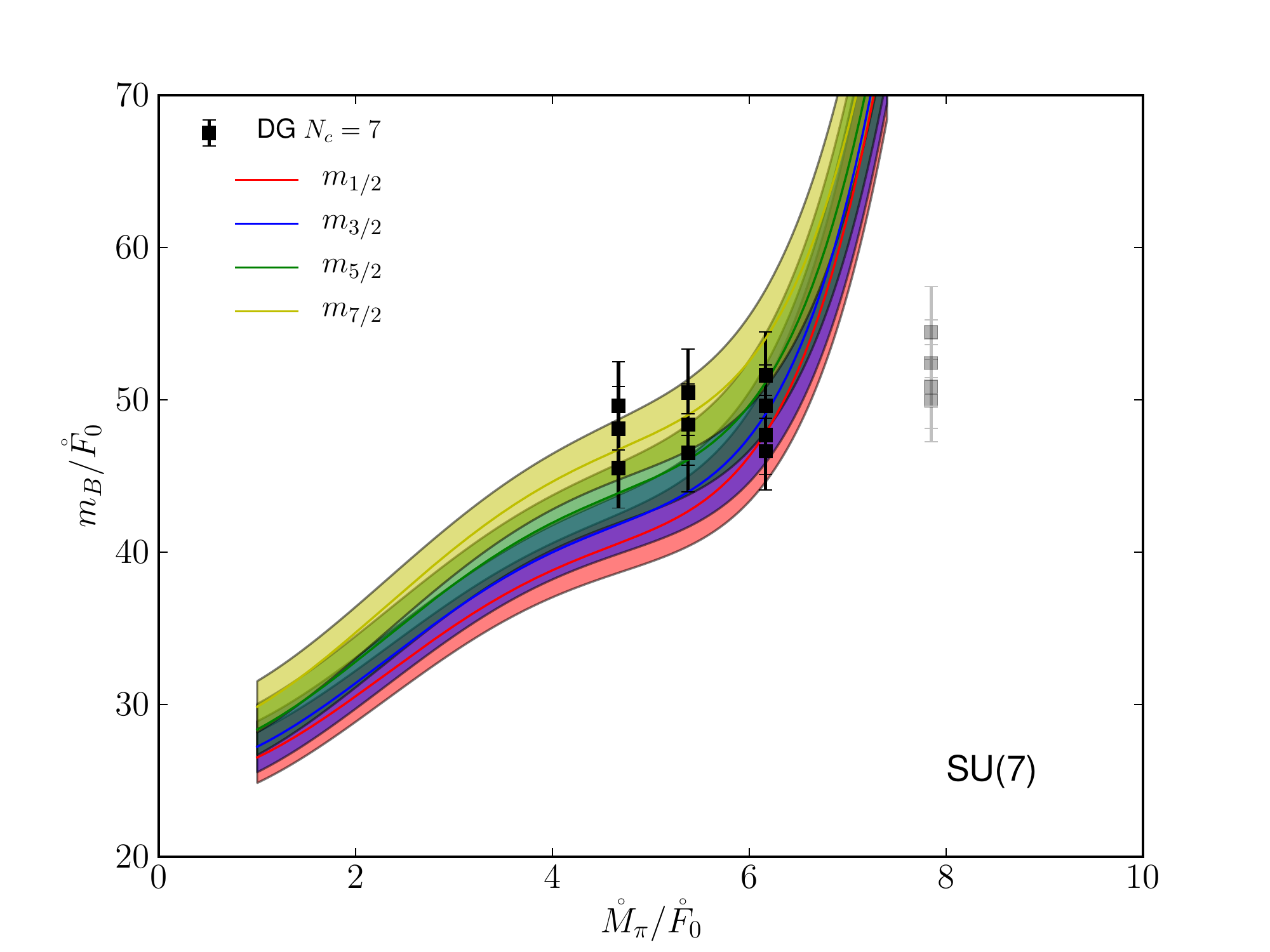}
\caption{Baryon masses for $N_c=3$, 5 and 7. Results in units of $\mathring{F}_0=\mathring{F}_\pi \sqrt{3/N_c}$. $N_c=3$  includes the results from PACS-CS~\cite{Aoki:2008sm}.  Shaded  points represent lattice points excluded in the fit which have an approximated pion mass of $M_\pi \gtrsim 700$ MeV.
}
\label{fig:Masses}
\end{figure}
\end{center}

As emphasized in Sec.~\ref{sec:BChPT}, the HF splittings  have a  better behaved low energy expansion than the masses themselves.  In fact they can be fitted up to   $M_\pi \lesssim 700$ MeV with natural magnitudes for the LECs.  The result is shown in Table~\ref{tab:results} as Fit I and in Fig.~\ref{fig:MassSplitting} as the dashed lines.
The LQCD results show a significant $M_\pi$ dependency of the HF splittings,  implying  that the corrections $\ord{M_\pi^2/N_c}$, determined by the loop contributions and the CT proportional to $\mu_2$, are important.
One finds that the minimum set of LECs needed to obtain a good fit to the HF splittings reduces to $C_{HF}$, $C_{HF1}$, $\mu_2$ and $z_1$. 
As shown by  Eqs.~\eqref{eq:dSigmaUV} and~\eqref{eq:dSigmaCT},  $C_{HF1}$ is the LEC of an  $O(1/N_c^2)$ CT,  while $\mu_2$ and $z_1$ are LECs of  $\ord{M_\pi^2}$ CTs.
In the fit one finds that the LECs $\mu_2$ and $z_1$ are highly correlated and thus cannot be determined using   the HF fit only. 
Setting one of them to vanish does not affect the $\chi^2$.  In  Fit I in Table~\ref{tab:results}   $z_1$ was set  to vanish.

The LEC $C_{HF1}$ gives the $1/N_c$ correction to the LO splitting determined by $C_{HF}$, and therefore some correlation between them is expected, as it is indeed the case: $C_{HF}$ changes significantly from the $\ord{\xi}$ fit to the $\ord{\xi^3}$ fit.
The LEC $z_1$ along with the rest of the LECs are most sensitive to the baryon masses, and are determined in the combined fit.
A detailed analysis of the fits shows that the LECs $m_2$, $m_3$, $w_1$, $w_2$, $C_{HF2}$ and $\mu_1$    have  marginal effects (due to correlations with the rest of the LECs),  and they are therefore be set to vanish.
\begin{table*}[htdp]
\caption{
LECs from fits to HF splittings (Fit I) and masses (Fits  II and III) at varying $N_c$ and a given order in the $\xi$-expansion $\ord{\xi^\nu}$. In the fits only data up to $M_\pi\sim 600$ MeV are included.
In Fits I and II,  $\mathring{g}_A=1.4$ and the renormalization scale is set at $\mu = 740$ MeV. In Fit III, $\mathring{g}_A=0$. 
The LECs are given in units of $\mathring{F}_0$ in the first set of rows.
The second set of rows  show the estimated values in physical units using $\mathring F_\pi = 90$ MeV. }
\begin{center}
\resizebox{\textwidth}{!}
{
\begin{tabular}{|c|c|c | c c c c c c c |}
\hline\hline
\multicolumn{10}{|c|}{LECs in units of $\mathring{F}_0$} \\\hline\hline
Fit & Order  & $\chi^2_{\rm DOF}$ 
    & $m_0$ & $m_{1}$ & $C_{HF}$ & $c_1$ & $C_{HF1}$ & $\mu_2$ & $z_1$ \\\hline
 I  & 3 & 0.70 & 0 & 0 &  0.39(9) & 0 & 6.7(3) & -0.040(4) & 0 \\\hline
\multirow{3}{*}{II} 
    & 1  & 1.48 & 6.0 (2) & -8.6 (7) & 2.86 (6) & 0.052 (2) & 0 & 0 & 0 \\
    & 2  & 1.07 & 5.7 (2) & -8.4 (6) & 0.96 (9) & 0.149 (3) & 6.3 (6) & 0 & 0 \\  
    & 3  & 0.89 & 4.0 (2) & -3.8 (7) & 0.61 (9) & 0.266 (5) & 6.1 (4) & 0.017 (3) & -0.0080 (4) \\\hline
III & 3  & 0.73 & 6.2 (2) & -9.2 (6) & 2.7 (3) &  0.050 (2) & 4.2 (7) & -0.004 (2) & -0.0040 (5) \\
\hline\hline
\multicolumn{10}{|c|}{LECs in physical units} \\\hline\hline
Fit & Order & $\chi^2_{\rm DOF}$ 
    & $m_0$ [MeV] & $m_{1}$ [MeV] & $C_{HF}$ [MeV] & $c_1$ (10$^{-3}$) [MeV$^{-1}$] &
    $C_{HF1}$ [MeV] & $\mu_2$  (10$^{-3}$) [MeV$^{-1}$] & $z_1$ (10$^{-6}$) [MeV$^{-2}$] \\\hline  
I  & 3 & 0.70 & 0 & 0 &  35(8) & 0 & 607 (31) & -0.42(4) & 0  \\\hline
\multirow{3}{*}{II} 
    & 1  & 1.48 & 543 (20) & -775 (60) & 257 (6) & 0.58 (3) & 0 & 0 & 0\\
    & 2  & 1.07 & 515 (20) & -752 (60) & 86 (8) & 1.66 (3) & 566 (60) & 0 & 0\\
    & 3  & 0.89 & 365 (20) & -345 (60) & 52 (7) & 2.96 (5) & 546 (40) & 0.20 (4) & -0.97 (4) \\\hline
III & 3  & 0.73 & 553 (20) & -827 (50) & 242 (22) & 0.50 (2) &  375 (60) & -0.04 (2) & -0.45 (6)\\\hline
\multicolumn{3}{|c}{Natural value} & $\sim$ 300 & $\sim$ & $\sim$ 300 & $\sim$ 10$^{-3}$ & $\sim$ & $\sim$ 10$^{-3}$  & $\sim$ 10$^{-6}$ 
\\  
\hline\hline
\end{tabular}
}
\end{center}
\label{tab:results}
\end{table*}
%

After the fit to the HF splittings, the combined fit including baryon masses is carried out. The results are shown in Table~\ref{tab:results} as Fit II and in Figs.~\ref{fig:MassSplitting} and~\ref{fig:Masses} as the solid lines. The errors shown in the Table represent the $68\%$ confidence interval and the bands are the corresponding Monte Carlo propagation of the LECs' errors.
 The values of the LECs $C_{HF}$ and $C_{HF1}$ previously determined by the HF fit to $\ord{\xi^3}$ change within errors, and the correlation between $\mu_2$ and $z_1$ is now eliminated and both LECs can be determined. The new result for the HF splittings is consistent with the ones of the Fit I  as shown in Fig.~\ref{fig:MassSplitting}, where the dashed lines of Fit I and full lines of Fit II fall inside the error bands. The LEC $z_1$ is needed in order to obtain a good description of the masses. This parameter is the leading in $1/N_c$ term contributing to the wave function renormalization and the value obtained here perfectly agrees with the one obtained for $N_c=3$   in Ref.~\cite{CalleCordon:2012xz}, where also the axial couplings were simultaneously analyzed.

In contrast to the HF splittings, the spin-flavor singlet component is naturally big,  $\ord{N_c}$, and receives corrections of the same order in $N_c$ but higher order in $M_\pi$. Thus,  the convergence of the low-energy expansion is poorer, with the   situation  increasingly worsening as $N_c$ increases. In contrast to the HF splittings,  the  combined fit  only has a range of validity at most up to  $M_\pi\sim 600$ MeV. For this reason the fits only include results with $M_\pi < 600$ MeV.
The LEC $m_{1}$ represents a $1/N_c$ correction to the term proportional to $m_0$,  and although  there is some  correlation among them, $m_{1}$ is very important. This is shown by Fig.~\ref{fig:Masses}, which   gives some evidence that the spin-flavor singlet part of the masses deviates  from the simple linear behavior in $N_c$.

Evidently  it will take a more extensive set of inputs at varying $N_c$ to fix all the LECs, with more results in the lower range of $M_\pi$ and possibly also larger values of $N_c$ than presently available. This obviously represents a difficult challenge at this time.
In order to study the stability of the low-energy expansion,  combined fits   were performed at different orders in  $\xi$.  The pattern of convergence of the spin-singlet LECs is stable. The pion mass dependence of the HF splitting only appears at   $\ord{\xi^3}$, through $\mu_2$, and therefore at lower orders one can only obtain a rough description. Clearly,  $C_{HF}$ is very sensitive to the  order of the expansion,  because it is strongly correlated with $C_{HF1}$.  

A comparison with the results for LECs obtained for $N_c=3$ in Ref.~\cite{CalleCordon:2012xz} requires the following identifications, where  on the left is the result of that fit and on the right the result of the present analysis: $m_0 \rightarrow m_0+\frac{1}{3}m_{1}$, $C_{HF}\rightarrow C_{HF}+\frac{1}{3} C_{HF1}$, $c_1\rightarrow c_1+\frac{1}{3} \mu_1$, $\mu_2\rightarrow\mu_2$ and $z_1\rightarrow z_1$. The respective comparisons are as follows: $250(30)$ MeV vs $310(15)$ MeV, $234(15)$ MeV vs $300(36)$  MeV, 0.00296(5) vs 0.00235(4) ($\mu_1$ set to zero), $0.20(4)\times 10^{-3} ~{\rm MeV}^{-1}$ vs $-7.3(6)\times 10^{-3} ~{\rm MeV}^{-1}$, and $-9.7(4)\times 10^{-7} ~{\rm MeV}^{-2}$ vs $-8.9(2)\times 10^{-7} ~{\rm MeV}^{-2}$.  These results  are in reasonable agreement,  taking into account that the LECs obtained in Ref.~\cite{CalleCordon:2012xz} were obtained from combined fits to masses and   the axial charges. Only $\mu_2$ is clearly in disagreement. Below the origin of the instability in the determination of $\mu_2$ will be discussed.

An additional test of consistency between  quenched and unquenched results is carried out by performing separate  fits at $N_c=3$ with only quenched or only unquenched LQCD results. The results of those fits are shown in  Table~\ref{tab:results-Nc3}: the first fit   only includes unquenched PACS-CS data,  the second only quenched DeGrand data, and finally   a third fit including both PACS-CS and DeGrand data.  Results  are shown in physical units using $\mathring F_\pi= 90$ MeV as the unit to transform.  Also shown  are the extrapolated nucleon and $\Delta$ masses to the physical point. 
\begin{table}[htdp]
\caption{
Independent fits to mass and HF splitting ratios of $N_c=3$ LQCD data sets. 
To convert to physical units we use $\mathring F_\pi= 90$ MeV
}
\begin{center}
\resizebox{\textwidth}{!}
{
\begin{tabular}{c|c|c|c|c|c|c|c|c}\hline\hline
LQCD Collaboration  & $\chi^2_{\rm DOF}$ & $m_0$ [MeV] & $C_{HF}$ [MeV] & $c_1$ (10$^{-3}$) [MeV$^{-1}$] & $\mu_2$  (10$^{-3}$) [MeV$^{-1}$] & $z_1$ (10$^{-6}$) [MeV$^{-2}$] & $m_N$ [MeV] &  $m_\Delta$ [MeV] \\\hline  
PACS-CS
& 0.47 & 253 (3) & 176 (8) & 2.5 (1) & -0.3 (1) & -0.7 (1) & 941 (2) & 1190 (10) \\\hline  
DeGrand
& 0.46 & 249 (4) & 184 (9) & 2.6 (2) & -0.09 (8) & -0.7 (1) & 939 (1) & 1203 (12)  \\\hline  
PACS-CS $+$ DeGrand 
& 0.67 & 255 (3) & 166 (7) & 2.6 (1) & -0.06 (5) & -0.76 (8) & 942 (2) & 1178 (8) \\
\hline\hline
\end{tabular}
}
\end{center}
\label{tab:results-Nc3}
\end{table}
 The independent fits   in Table~\ref{tab:results-Nc3} show  compatibility of the quenched and unquenched results at $N_c=3$. This lends a strong support to the approach used here for combining results at the level of masses in units of $\mathring F_\pi$.
One therefore concludes that,  given that quenched and full QCD are formally   different at sub leading order in $1/N_c$, the phenomenological approach followed here  shows that   such differences    are not clearly noticeably in the  present  analysis.   

\begin{center}
\begin{figure}[h!!!!]
\includegraphics[width=5. cm, height=5. cm,angle=0]{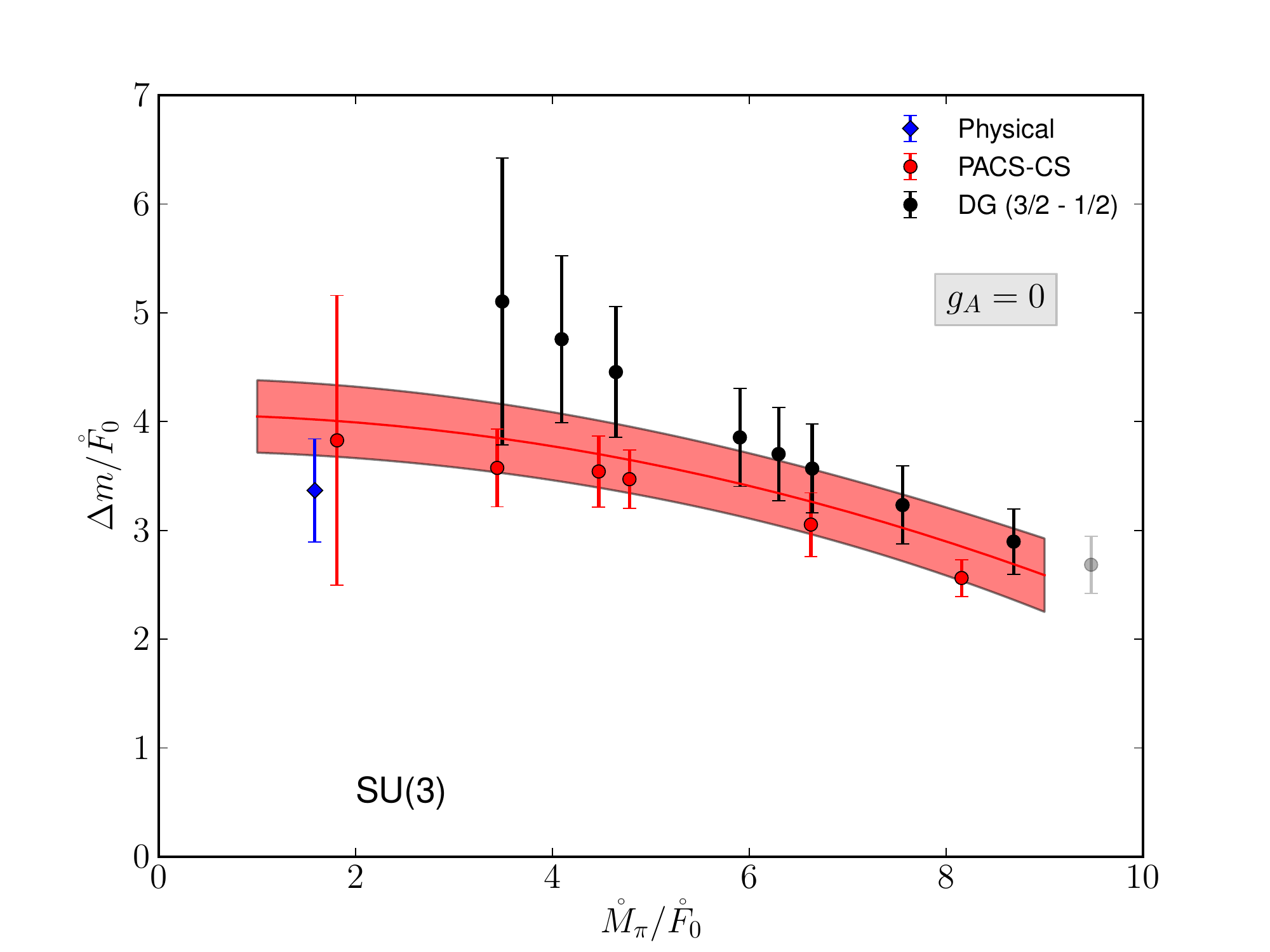}
\includegraphics[width=5. cm, height=5. cm,angle=0]{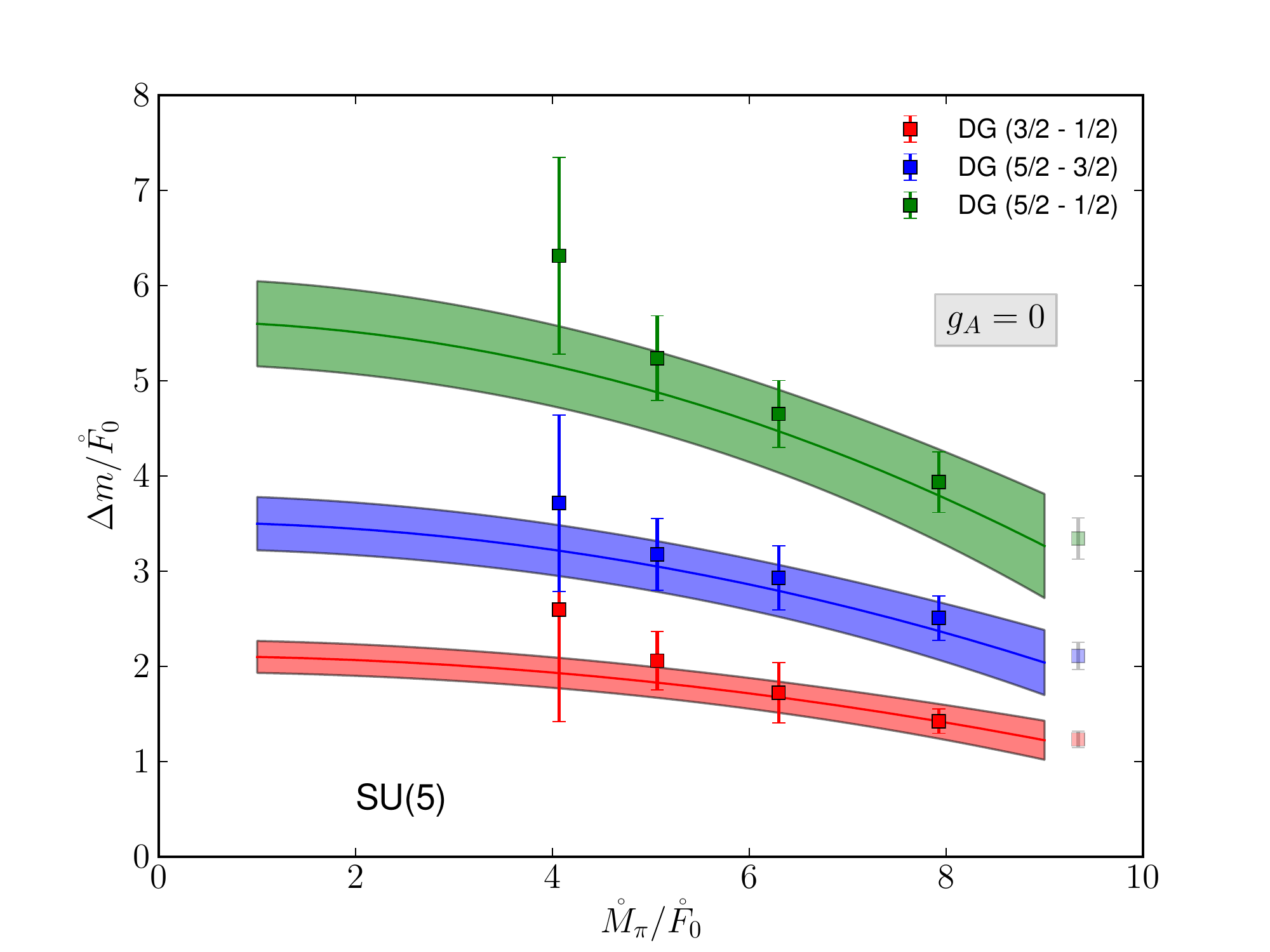}
\includegraphics[width=5. cm, height=5. cm,angle=0]{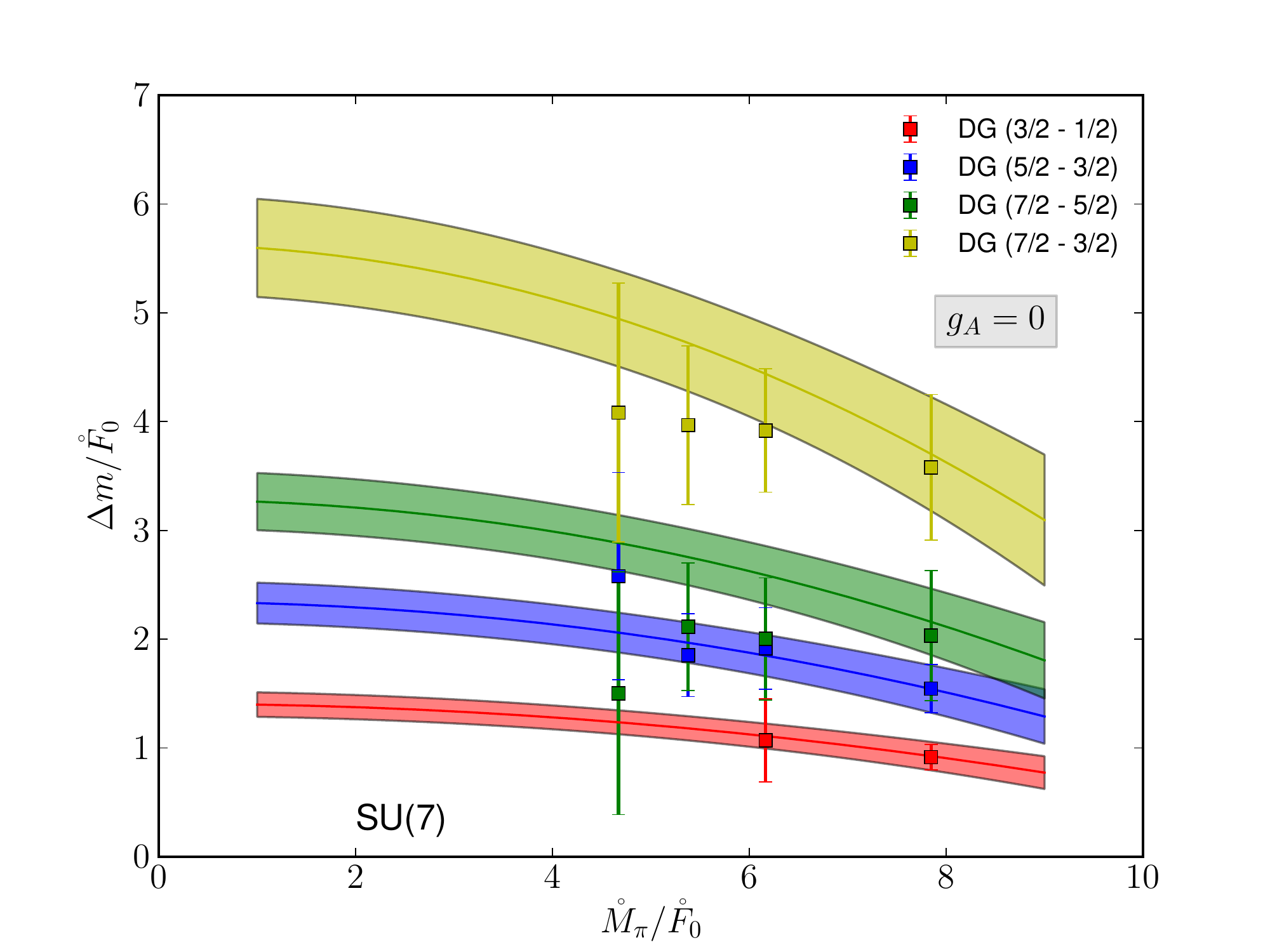}
\caption{HF splittings in the $N_c=3$, 5 and 7 multiplets obtained by setting $\mathring{g}_A=0$.  Results are in units of $\mathring F_0$. For $N_c=3$,   the  results of PACS-CS~\cite{Aoki:2008sm} are also included.
}
\label{fig:MassSplitting-noloop}
\vspace*{1.cm}
\includegraphics[width=5.cm, height=5cm,angle=0]{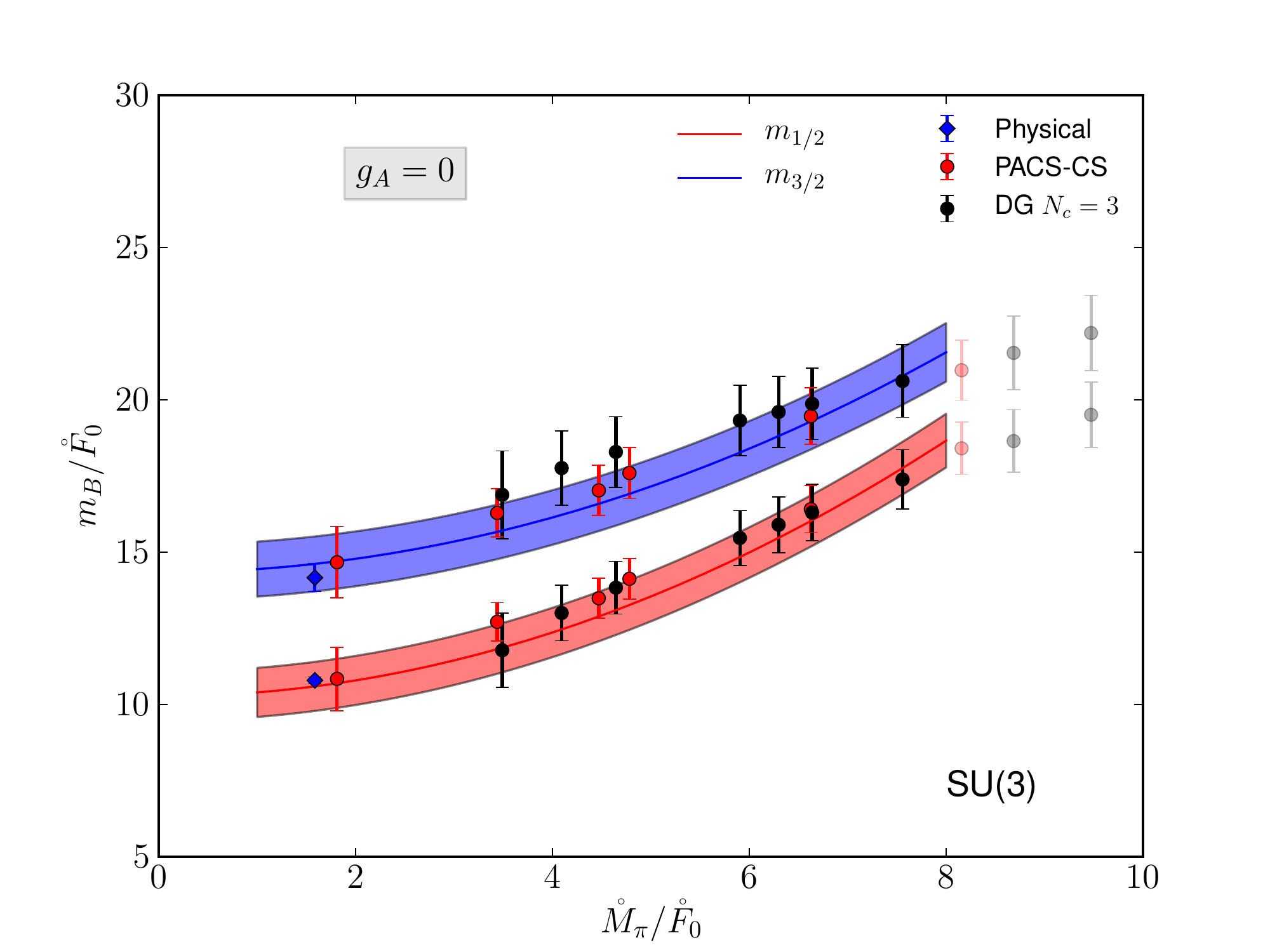}
\includegraphics[width=5.cm, height=5cm,angle=0]{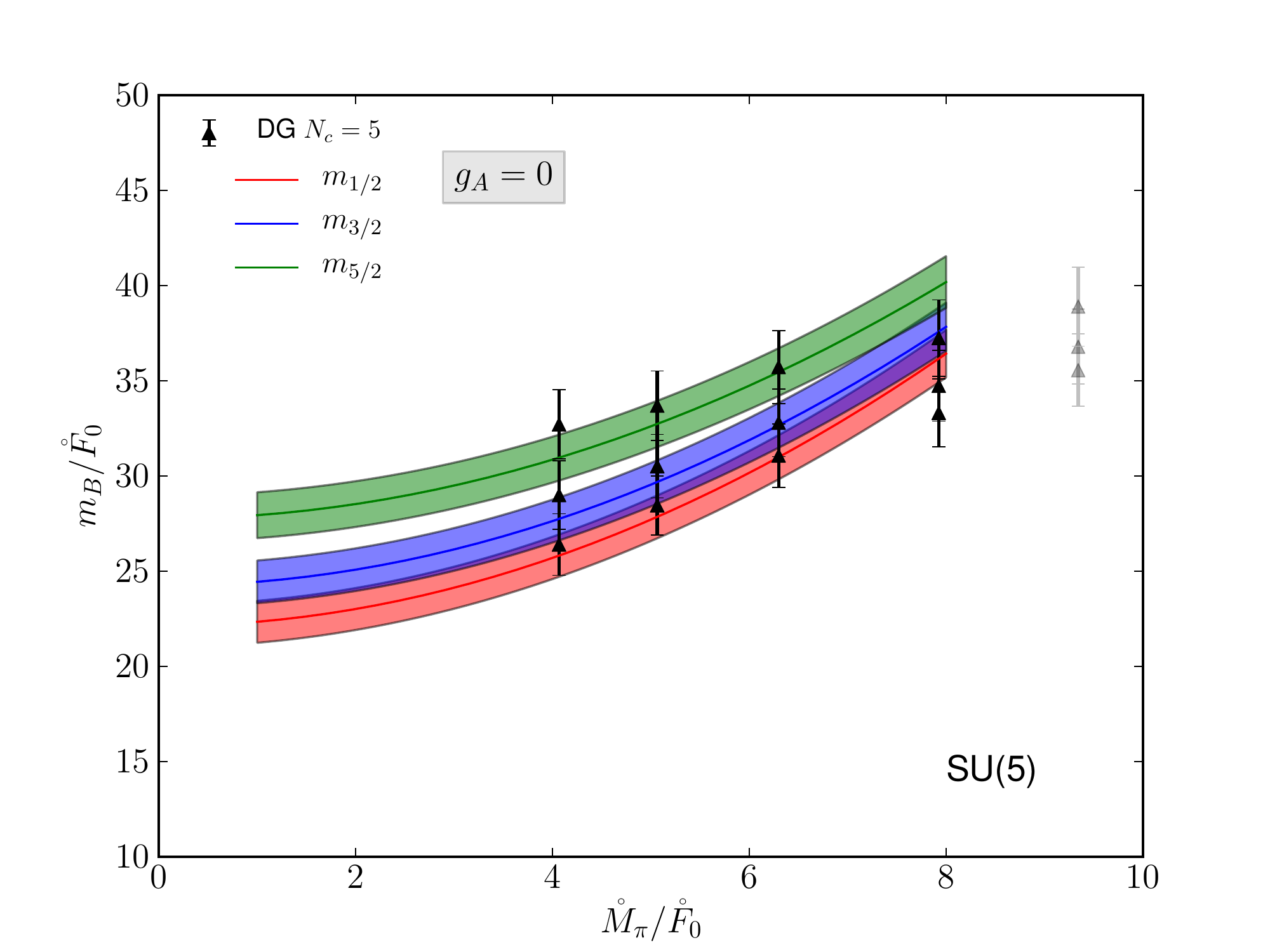}
\includegraphics[width=5.cm, height=5cm,angle=0]{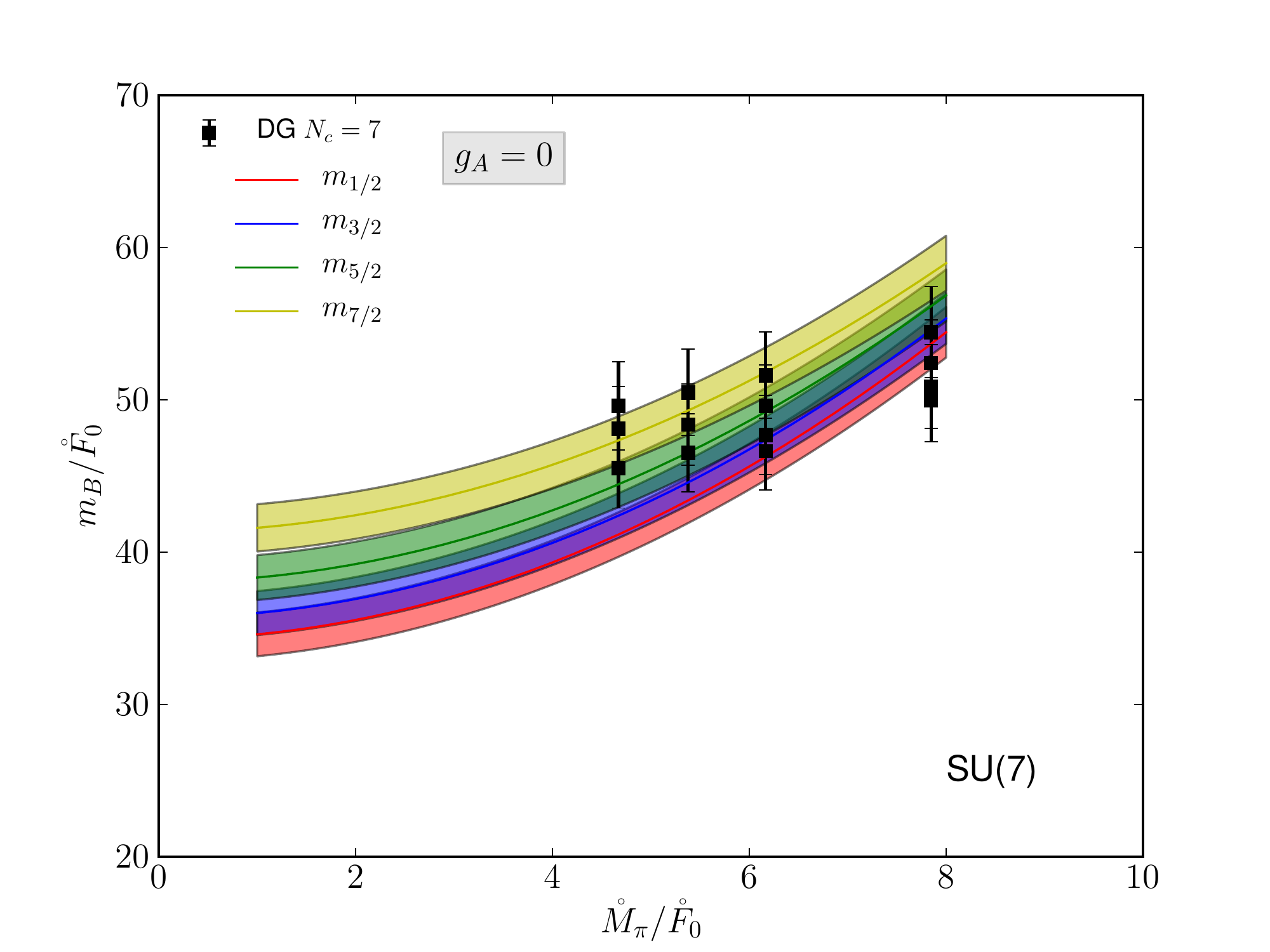}
\caption{Baryon masses   $N_c=3$, 5 and 7  obtained by  setting $\mathring{g}_A=0$. Results are in units of $\mathring F_0$. For $N_c=3$, the results of  PACS-CS~\cite{Aoki:2008sm} are also included.}
\label{fig:Masses-noloop}
\end{figure}
\end{center}

It is  important to test the one loop contributions in the description of the LQCD results. In order to expose them, a combined fit is carried out at $\ord{\xi^3}$ in which the coupling $\mathring{g}_A$ is set to vanish (Fit. III in Table~\ref{tab:results} and Figs.~\ref{fig:MassSplitting-noloop} and~\ref{fig:Masses-noloop}). As expected, the fit can be consistently carried out  for larger values of $M_\pi$. When a similar range of $M_\pi$ is used,  the LECs show larger error bars, which can be explained by the fact that in the case where $\mathring{g}_A=1.4$ there must be  important cancellations between  loop contributions and counter-terms leading to tighter error bars.
 The impact of the loop contributions can be seen in the very different chiral extrapolations obtained with and without the one loop contributions.  
Let us estimate the range in $M_\pi$ where the EFT seems to work. This can be estimated by taking  as the upper boundary of that range  the upper inflection points of the curves in Fig.~\ref{fig:Masses}. This gives a range approximately described by  $M_\pi/\mathring{F}_0< 6.25-0.25\,N_c$.

 Finally, the $1/N_c$ expansion, where the link  $1/N_c=\ord{p^2}$ is used, turns out to give poor fits, even when one removes the inputs at lower values of $M_\pi$ where it should not work. The only way it can give a reasonable fit is if the effects of the wave function renormalization are enormous. This is unrealistic.

The analysis leads to the following observations:
\begin{enumerate}
\item In the range of $M_\pi$  considered, the chiral loop contributions are very important, driving the curvature observed in the HF splittings, and giving extrapolations of the baryon masses to small pion mass which are very different than in the case where the loop contributions are excluded. This effect increases with $N_c$.
\item The fact that the range  of applicability   in $M_\pi$ of the EFT  diminishes with increasing $N_c$, means that LQCD results for $N_c=5$ and 7 at smaller quark masses than the ones presented in this work would be necessary for a more significant analysis. In addition, 
a proper analysis requires unquenched results.

\item The $N_c=7$ inputs have played a lesser role in the results of the analysis because they are mostly located in a range of pion masses where the convergence of the EFT is  unwarranted. Only results at lowest three values of $M_\pi$ could be included in the fits.

\item  The naturalness of the results is reflected in the fact that in units of GeV the LECs are all of order one, as one would expect from a well behaved expansion. 

\item The two new effects that have become accessible with the LQCD results are the sub-leading in $1/N_c$ terms in the spin-flavor singlet component of the baryon masses and in the HF splittings.   The LECs associated with those effects, namely $m_1$ and $C_{HF1}$ 
respectively, have been determined. The latter one is   larger than the expected natural size, but it must be noticed that it is very strongly correlated with $C_{HF}$, which is the leading $HF$ LEC. 

\item    
The LQCD results show clearly that the HF splittings   decrease with increasing quark mass in approximately the same proportion for all values of $N_c$. This behavior has contributions from the chiral loop as well as the LEC $\mu_2$. There is significant curvature as shown if Fig. \ref{fig:MassSplitting} due to the loop contributions. This makes the determination of $\mu_2$ rather uncertain. In fact, its value is sensitive to the range of $M_\pi$ considered in the fit. It is possible that one could use the  stability of $\mu_2$ to set the range in $M_\pi$ where the fit to the HF splittings can be safely done with the EFT.

\end{enumerate}

\section{Conclusions} \label{sec:Conclusions}

Understanding the $N_c$ dependence of gauge theories is a fundamental theoretical problem, which also has profound phenomenological consequences for QCD.
In fact, QCD phenomenology in both mesons and baryons indicates that  an expansion in powers of $1/N_c$  seems to work for most observables. The actual study of QCD at different values of $N_c$ is  essential for confirming that observation.
LQCD provides the only present means for those studies  beyond $N_c=3$. These studies are essential for quantifying the sub-leading in $1/N_c$ effects.  
In the case of light baryons studied in this work,  the aim has been  to  use the LQCD results up to $N_c=7$ in order to elucidate sub-leading in $1/N_c$ effects as described by a low energy EFT.       The study proved to be
fruitful, showing consistency with previous results obtained by analyzing the $N_c=3$ LQCD results, and in addition  it permitted the  determination of  two  sub-leading effects, namely the sub-leading correction to the spin-flavor singlet component of the baryon masses, which is entirely given by the LEC $m_1$, and the sub-leading corrections to the HF splittings which stem from loop contributions and the LEC $C_{HF1}$. Other sub-leading effects which are in principle present could not be determined because of the still significant error bars in the LQCD results.

It is observed that the different LQCD results for the HF splittings are not entirely compatible. 
Since HF effects are dominated by short distance contributions, they may be very sensitive to lattice
artifacts, and might require careful continuum extrapolations.
 It is clearly desirable to have more accurate HF results, in particular because the HF effects are better behaved in the EFT than the masses themselves.

It is clear that the problem of chiral extrapolations of baryon masses (not of HF splittings)  becomes more severe  as a consequence of the shrinking domain of the low energy expansion with 
increasing $N_c$. In order to improve the study of this work, it will be necessary to have $N_c>3$  LQCD results  for smaller quark masses than the ones analyzed. 
For a fully consistent study the next step should   include dynamical quarks, in particular because of the problem of exceptional configurations which arises in quenched QCD as the quark mass decreases. This task is clearly very challenging, but we believe that it is not impossible,
merely expensive.
It is one that will give important additional insights into the   $1/N_c$ expansion for light baryons.  Finally, the extension to three flavors based on the recent results \cite{DeGrand:2013nna} can be already analyzed in the same way as the case of two flavor presented here.

\begin{acknowledgments}
T.~D. thanks JLab for its hospitality, where this project was initiated.
The conversion of the MILC code to arbitrary number of colors was done by T.~D. with Y.~Shamir and B.~Svetitsky.
This work was supported in part by the U.~S. Department of Energy under grant DE-FG02-04ER41290 (T.~D.),
by DOE Contract No. DE-AC05-06OR23177 under which JSA operates the Thomas Jefferson National Accelerator Facility (A.C.C and J.~L.~G.), and by the National Science Foundation (USA) through grant PHY-0855789 and PHY-1307413 (J.~L.~G.). 
A.C.C also thanks the financial support from the EU-Research Infrastructure Integrating Activity, \lq\lq Study of Strongly Interacting Matter\rq\rq (HadronPhysics2, grant n.227431) under the Seventh Framework Program of the EU.
\end{acknowledgments}

\newpage

\appendix

\section{UV finite parts of the self-energy}

The one-loop finite contributions to the  self-energy relevant for the masses at $\ord{\xi^3}$ are as follows:
{
\footnotesize
\bea
\!\!\!\!\!\!\!\!\!\!\!\!\!\!&&\delta \Sigma_{\rm finite}(\p=0)(S)=
\frac{
{\mathring{g}_A}^2({N_c}+4)}{128 \pi ^2 {\mathring F_0}^2} \nonumber\\
 &\times&\left\{\frac{2
   \left(3(2 S-1)
   {A_1}(S-1)-3 (2 S+3)
   {A_1}(S)+8 \,{C_{HF}}^3
   (2 S+1) (5 S (S+1)+3)
   \left(4-3 \log
   \left(\frac{{M_\pi}}{\mu
   }\right)\right)\right)}{3
   {N_c}^3 (2 S+1)}\right.\nonumber\\
   &+&\pi  \left(\frac{4\, {C_{HF}}^2
   \left((2 S-1) S^2
   {R}(S-1)+(S+1)^2 (2
   S+3)
   {R}(S)\right)}{{
   N_c}^3 (2
   S+1)}-{M_\pi}^3+\frac{{M_\pi}^2 ((1-2 S)
   {R}(S-1)-(2 S+3)
   {R}(S))}{ {N_c}(2
   S+1)}\right)\nonumber\\
   &+&\left.\frac{3\,
   {C_{HF}} {M_\pi}^2
   \left(6 \log
   \left(\frac{{M_\pi}}{\mu
   }\right)-7\right)}{{N_c
   }}\right\}
\eea

\bea
\delta Z_{\rm finite}(S)&=&
\frac{3\,{\mathring{g}_A}^2({N_c}+4)}{64\pi^2 {\mathring F_0}^2}
   \left\{\frac{1}{{N_c}^2}\left(\frac{{C_{HF}
   } (S (2 S-1)
   {A_2}(S-1)+(S+1) (2
   S+3) {A_2}(S))}{ (2
   S+1)}\right.\right.\nonumber\\
   &+&\left.\left. 2\,
   {C_{HF}}^2 (2 S (S+1)+3)
   \left(\log
   \left(\frac{{M_\pi}}{\mu
   }\right)-1\right)+\frac{\pi\,
    {C_{HF}} (S (2 S-1)
   {R}(S-1)-(S+1) (2
   S+3) {R}(S))}{2 (2
   S+1)}\right)\right.\nonumber\\
   &+&\left.\frac{1
   }{8} {M_\pi}^2
   \left(3-6 \log
   \left(\frac{{M_\pi}}{\mu
   }\right)\right)\right \}
   \eea
}

where :
\bea
R(S)&=&\sqrt{N_c^2 M_\pi^2-4(S+1)^2 \,C_{HF}^2}\nonumber\\
A_1(S)&=&\left(-N_c^2 M_\pi^2+4(S+1)^2\, C_{HF}^2\right)^{\frac 32} \;{\rm arctanh}\left(\frac{2(S+1)\, C_{HF}}{\sqrt{-N_c^2 M_\pi^2+4(S+1)^2\, C_{HF}^2}}\right)\nonumber\\
A_2(S)&=&\frac{A_1(S)}{-N_c^2 M_\pi^2+4(S+1)^2 \, C_{HF}^2}
 \eea
 
 The terms in  Eqns. (A1) and (A2) involving $R$, $A_1$ and $A_2$ are due to the contributions of the baryons with spins $S\pm 1$ in the loop. 

\newpage
\bibliography{Refs}

\begin{thebibliography}{45}
\expandafter\ifx\csname natexlab\endcsname\relax\def\natexlab#1{#1}\fi
\expandafter\ifx\csname bibnamefont\endcsname\relax
  \def\bibnamefont#1{#1}\fi
\expandafter\ifx\csname bibfnamefont\endcsname\relax
  \def\bibfnamefont#1{#1}\fi
\expandafter\ifx\csname citenamefont\endcsname\relax
  \def\citenamefont#1{#1}\fi
\expandafter\ifx\csname url\endcsname\relax
  \def\url#1{\texttt{#1}}\fi
\expandafter\ifx\csname urlprefix\endcsname\relax\def\urlprefix{URL }\fi
\providecommand{\bibinfo}[2]{#2}
\providecommand{\eprint}[2][]{\url{#2}}

\bibitem[{\citenamefont{Teper}(1997)}]{Teper:1997tq}
\bibinfo{author}{\bibfnamefont{M.}~\bibnamefont{Teper}},
  \bibinfo{journal}{Phys.Lett.} \textbf{\bibinfo{volume}{B397}},
  \bibinfo{pages}{223} (\bibinfo{year}{1997}), \eprint{hep-lat/9701003}.

\bibitem[{\citenamefont{Lucini and Teper}(2001)}]{Lucini:2001ej}
\bibinfo{author}{\bibfnamefont{B.}~\bibnamefont{Lucini}} \bibnamefont{and}
  \bibinfo{author}{\bibfnamefont{M.}~\bibnamefont{Teper}},
  \bibinfo{journal}{JHEP} \textbf{\bibinfo{volume}{0106}}, \bibinfo{pages}{050}
  (\bibinfo{year}{2001}), \eprint{hep-lat/0103027}.

\bibitem[{\citenamefont{Del~Debbio et~al.}(2002)\citenamefont{Del~Debbio,
  Panagopoulos, Rossi, and Vicari}}]{DelDebbio:2001sj}
\bibinfo{author}{\bibfnamefont{L.}~\bibnamefont{Del~Debbio}},
  \bibinfo{author}{\bibfnamefont{H.}~\bibnamefont{Panagopoulos}},
  \bibinfo{author}{\bibfnamefont{P.}~\bibnamefont{Rossi}}, \bibnamefont{and}
  \bibinfo{author}{\bibfnamefont{E.}~\bibnamefont{Vicari}},
  \bibinfo{journal}{JHEP} \textbf{\bibinfo{volume}{0201}}, \bibinfo{pages}{009}
  (\bibinfo{year}{2002}), \eprint{hep-th/0111090}.

\bibitem[{\citenamefont{Allton et~al.}(2008)\citenamefont{Allton, Teper, and
  Trivini}}]{Allton:2008ty}
\bibinfo{author}{\bibfnamefont{C.}~\bibnamefont{Allton}},
  \bibinfo{author}{\bibfnamefont{M.}~\bibnamefont{Teper}}, \bibnamefont{and}
  \bibinfo{author}{\bibfnamefont{A.}~\bibnamefont{Trivini}},
  \bibinfo{journal}{JHEP} \textbf{\bibinfo{volume}{0807}}, \bibinfo{pages}{021}
  (\bibinfo{year}{2008}), \eprint{0803.1092}.

\bibitem[{\citenamefont{Lucini et~al.}(2004)\citenamefont{Lucini, Teper, and
  Wenger}}]{Lucini:2004my}
\bibinfo{author}{\bibfnamefont{B.}~\bibnamefont{Lucini}},
  \bibinfo{author}{\bibfnamefont{M.}~\bibnamefont{Teper}}, \bibnamefont{and}
  \bibinfo{author}{\bibfnamefont{U.}~\bibnamefont{Wenger}},
  \bibinfo{journal}{JHEP} \textbf{\bibinfo{volume}{0406}}, \bibinfo{pages}{012}
  (\bibinfo{year}{2004}), \eprint{hep-lat/0404008}.

\bibitem[{\citenamefont{Lucini et~al.}(2010)\citenamefont{Lucini, Rago, and
  Rinaldi}}]{Lucini:2010nv}
\bibinfo{author}{\bibfnamefont{B.}~\bibnamefont{Lucini}},
  \bibinfo{author}{\bibfnamefont{A.}~\bibnamefont{Rago}}, \bibnamefont{and}
  \bibinfo{author}{\bibfnamefont{E.}~\bibnamefont{Rinaldi}},
  \bibinfo{journal}{JHEP} \textbf{\bibinfo{volume}{1008}}, \bibinfo{pages}{119}
  (\bibinfo{year}{2010}), \eprint{1007.3879}.

\bibitem[{\citenamefont{Del~Debbio et~al.}(2008)\citenamefont{Del~Debbio,
  Lucini, Patella, and Pica}}]{DelDebbio:2007wk}
\bibinfo{author}{\bibfnamefont{L.}~\bibnamefont{Del~Debbio}},
  \bibinfo{author}{\bibfnamefont{B.}~\bibnamefont{Lucini}},
  \bibinfo{author}{\bibfnamefont{A.}~\bibnamefont{Patella}}, \bibnamefont{and}
  \bibinfo{author}{\bibfnamefont{C.}~\bibnamefont{Pica}},
  \bibinfo{journal}{JHEP} \textbf{\bibinfo{volume}{0803}}, \bibinfo{pages}{062}
  (\bibinfo{year}{2008}), \eprint{0712.3036}.

\bibitem[{\citenamefont{Bali and Bursa}(2008)}]{Bali:2008an}
\bibinfo{author}{\bibfnamefont{G.~S.} \bibnamefont{Bali}} \bibnamefont{and}
  \bibinfo{author}{\bibfnamefont{F.}~\bibnamefont{Bursa}},
  \bibinfo{journal}{JHEP} \textbf{\bibinfo{volume}{0809}}, \bibinfo{pages}{110}
  (\bibinfo{year}{2008}), \eprint{0806.2278}.

\bibitem[{\citenamefont{Bali et~al.}(2013)\citenamefont{Bali, Bursa,
  Castagnini, Collins, Del~Debbio et~al.}}]{Bali:2013kia}
\bibinfo{author}{\bibfnamefont{G.~S.} \bibnamefont{Bali}},
  \bibinfo{author}{\bibfnamefont{F.}~\bibnamefont{Bursa}},
  \bibinfo{author}{\bibfnamefont{L.}~\bibnamefont{Castagnini}},
  \bibinfo{author}{\bibfnamefont{S.}~\bibnamefont{Collins}},
  \bibinfo{author}{\bibfnamefont{L.}~\bibnamefont{Del~Debbio}},
  \bibnamefont{et~al.}, \bibinfo{journal}{JHEP}
  \textbf{\bibinfo{volume}{1306}}, \bibinfo{pages}{071} (\bibinfo{year}{2013}),
  \eprint{1304.4437}.

\bibitem[{\citenamefont{Narayanan and Neuberger}(2005)}]{Narayanan:2005gh}
\bibinfo{author}{\bibfnamefont{R.}~\bibnamefont{Narayanan}} \bibnamefont{and}
  \bibinfo{author}{\bibfnamefont{H.}~\bibnamefont{Neuberger}},
  \bibinfo{journal}{Phys.Lett.} \textbf{\bibinfo{volume}{B616}},
  \bibinfo{pages}{76} (\bibinfo{year}{2005}), \eprint{hep-lat/0503033}.

\bibitem[{\citenamefont{Hietanen et~al.}(2009)\citenamefont{Hietanen,
  Narayanan, Patel, and Prays}}]{Hietanen:2009tu}
\bibinfo{author}{\bibfnamefont{A.}~\bibnamefont{Hietanen}},
  \bibinfo{author}{\bibfnamefont{R.}~\bibnamefont{Narayanan}},
  \bibinfo{author}{\bibfnamefont{R.}~\bibnamefont{Patel}}, \bibnamefont{and}
  \bibinfo{author}{\bibfnamefont{C.}~\bibnamefont{Prays}},
  \bibinfo{journal}{Phys.Lett.} \textbf{\bibinfo{volume}{B674}},
  \bibinfo{pages}{80} (\bibinfo{year}{2009}), \eprint{0901.3752}.

\bibitem[{\citenamefont{DeGrand}(2012)}]{DeGrand:2012hd}
\bibinfo{author}{\bibfnamefont{T.}~\bibnamefont{DeGrand}},
  \bibinfo{journal}{Phys.Rev.} \textbf{\bibinfo{volume}{D86}},
  \bibinfo{pages}{034508} (\bibinfo{year}{2012}), \eprint{1205.0235}.

\bibitem[{\citenamefont{Vicari and Panagopoulos}(2009)}]{Vicari:2008jw}
\bibinfo{author}{\bibfnamefont{E.}~\bibnamefont{Vicari}} \bibnamefont{and}
  \bibinfo{author}{\bibfnamefont{H.}~\bibnamefont{Panagopoulos}},
  \bibinfo{journal}{Phys.Rept.} \textbf{\bibinfo{volume}{470}},
  \bibinfo{pages}{93} (\bibinfo{year}{2009}), \eprint{0803.1593}.

\bibitem[{\citenamefont{Lucini and Panero}(2013)}]{Lucini:2012gg}
\bibinfo{author}{\bibfnamefont{B.}~\bibnamefont{Lucini}} \bibnamefont{and}
  \bibinfo{author}{\bibfnamefont{M.}~\bibnamefont{Panero}},
  \bibinfo{journal}{Phys.Rept.} \textbf{\bibinfo{volume}{526}},
  \bibinfo{pages}{93} (\bibinfo{year}{2013}), \eprint{1210.4997}.

\bibitem[{\citenamefont{Jenkins}(1996)}]{Jenkins:1995gc}
\bibinfo{author}{\bibfnamefont{E.~E.} \bibnamefont{Jenkins}},
  \bibinfo{journal}{Phys.Rev.} \textbf{\bibinfo{volume}{D53}},
  \bibinfo{pages}{2625} (\bibinfo{year}{1996}), \eprint{hep-ph/9509433}.

\bibitem[{\citenamefont{Flores-Mendieta
  et~al.}(1998)\citenamefont{Flores-Mendieta, Jenkins, and
  Manohar}}]{FloresMendieta:1998ii}
\bibinfo{author}{\bibfnamefont{R.}~\bibnamefont{Flores-Mendieta}},
  \bibinfo{author}{\bibfnamefont{E.~E.} \bibnamefont{Jenkins}},
  \bibnamefont{and} \bibinfo{author}{\bibfnamefont{A.~V.}
  \bibnamefont{Manohar}}, \bibinfo{journal}{Phys.Rev.}
  \textbf{\bibinfo{volume}{D58}}, \bibinfo{pages}{094028}
  (\bibinfo{year}{1998}), \eprint{hep-ph/9805416}.

\bibitem[{\citenamefont{Flores-Mendieta and
  Hofmann}(2006)}]{FloresMendieta:2006ei}
\bibinfo{author}{\bibfnamefont{R.}~\bibnamefont{Flores-Mendieta}}
  \bibnamefont{and} \bibinfo{author}{\bibfnamefont{C.~P.}
  \bibnamefont{Hofmann}}, \bibinfo{journal}{Phys.Rev.}
  \textbf{\bibinfo{volume}{D74}}, \bibinfo{pages}{094001}
  (\bibinfo{year}{2006}), \eprint{hep-ph/0609120}.

\bibitem[{\citenamefont{Calle~Cordon and
  Goity}(2013{\natexlab{a}})}]{CalleCordon:2012xz}
\bibinfo{author}{\bibfnamefont{A.}~\bibnamefont{Calle~Cordon}}
  \bibnamefont{and} \bibinfo{author}{\bibfnamefont{J.~L.} \bibnamefont{Goity}},
  \bibinfo{journal}{Phys.Rev.} \textbf{\bibinfo{volume}{D87}},
  \bibinfo{pages}{016019} (\bibinfo{year}{2013}{\natexlab{a}}),
  \eprint{1210.2364}.

\bibitem[{\citenamefont{Calle~Cordon and Goity}(2012)}]{Cordon:2012nm}
\bibinfo{author}{\bibfnamefont{A.}~\bibnamefont{Calle~Cordon}}
  \bibnamefont{and} \bibinfo{author}{\bibfnamefont{J.~L.} \bibnamefont{Goity}},
  \bibinfo{journal}{PoS} \textbf{\bibinfo{volume}{QNP2012}},
  \bibinfo{pages}{136} (\bibinfo{year}{2012}), \eprint{1209.0030}.

\bibitem[{\citenamefont{Calle~Cordon and
  Goity}(2013{\natexlab{b}})}]{Cordon:2013era}
\bibinfo{author}{\bibfnamefont{A.}~\bibnamefont{Calle~Cordon}}
  \bibnamefont{and} \bibinfo{author}{\bibfnamefont{J.~L.} \bibnamefont{Goity}}
  (\bibinfo{year}{2013}{\natexlab{b}}), \eprint{1303.2126}.

\bibitem[{\citenamefont{Jenkins and
  Manohar}(1991{\natexlab{a}})}]{Jenkins:1990jv}
\bibinfo{author}{\bibfnamefont{E.~E.} \bibnamefont{Jenkins}} \bibnamefont{and}
  \bibinfo{author}{\bibfnamefont{A.~V.} \bibnamefont{Manohar}},
  \bibinfo{journal}{Phys.Lett.} \textbf{\bibinfo{volume}{B255}},
  \bibinfo{pages}{558} (\bibinfo{year}{1991}{\natexlab{a}}).

\bibitem[{\citenamefont{Jenkins and
  Manohar}(1991{\natexlab{b}})}]{Jenkins:1991ne}
\bibinfo{author}{\bibfnamefont{E.~E.} \bibnamefont{Jenkins}} \bibnamefont{and}
  \bibinfo{author}{\bibfnamefont{A.~V.} \bibnamefont{Manohar}},
  \bibinfo{journal}{Phys.Lett.} \textbf{\bibinfo{volume}{B259}},
  \bibinfo{pages}{353} (\bibinfo{year}{1991}{\natexlab{b}}).

\bibitem[{\citenamefont{Dashen and
  Manohar}(1993{\natexlab{a}})}]{Dashen:1993ac}
\bibinfo{author}{\bibfnamefont{R.~F.} \bibnamefont{Dashen}} \bibnamefont{and}
  \bibinfo{author}{\bibfnamefont{A.~V.} \bibnamefont{Manohar}},
  \bibinfo{journal}{Phys.Lett.} \textbf{\bibinfo{volume}{B315}},
  \bibinfo{pages}{438} (\bibinfo{year}{1993}{\natexlab{a}}),
  \eprint{hep-ph/9307242}.

\bibitem[{\citenamefont{Dashen and
  Manohar}(1993{\natexlab{b}})}]{Dashen:1993as}
\bibinfo{author}{\bibfnamefont{R.~F.} \bibnamefont{Dashen}} \bibnamefont{and}
  \bibinfo{author}{\bibfnamefont{A.~V.} \bibnamefont{Manohar}},
  \bibinfo{journal}{Phys.Lett.} \textbf{\bibinfo{volume}{B315}},
  \bibinfo{pages}{425} (\bibinfo{year}{1993}{\natexlab{b}}),
  \eprint{hep-ph/9307241}.

\bibitem[{\citenamefont{Flores-Mendieta
  et~al.}(2012)\citenamefont{Flores-Mendieta, Hernandez-Ruiz, and
  Hofmann}}]{FloresMendieta:2012dn}
\bibinfo{author}{\bibfnamefont{R.}~\bibnamefont{Flores-Mendieta}},
  \bibinfo{author}{\bibfnamefont{M.~A.} \bibnamefont{Hernandez-Ruiz}},
  \bibnamefont{and} \bibinfo{author}{\bibfnamefont{C.~P.}
  \bibnamefont{Hofmann}}, \bibinfo{journal}{Phys.Rev.}
  \textbf{\bibinfo{volume}{D86}}, \bibinfo{pages}{094041}
  (\bibinfo{year}{2012}), \eprint{1210.8445}.

\bibitem[{\citenamefont{DeGrand}(2014)}]{DeGrand:2013nna}
\bibinfo{author}{\bibfnamefont{T.}~\bibnamefont{DeGrand}},
  \bibinfo{journal}{Phys.Rev.} \textbf{\bibinfo{volume}{D89}},
  \bibinfo{pages}{014506} (\bibinfo{year}{2014}), \eprint{1308.4114}.

\bibitem[{\citenamefont{Gervais and
  Sakita}(1984{\natexlab{a}})}]{Gervais:1983wq}
\bibinfo{author}{\bibfnamefont{J.-L.} \bibnamefont{Gervais}} \bibnamefont{and}
  \bibinfo{author}{\bibfnamefont{B.}~\bibnamefont{Sakita}},
  \bibinfo{journal}{Phys.Rev.Lett.} \textbf{\bibinfo{volume}{52}},
  \bibinfo{pages}{87} (\bibinfo{year}{1984}{\natexlab{a}}).

\bibitem[{\citenamefont{Gervais and
  Sakita}(1984{\natexlab{b}})}]{Gervais:1984rc}
\bibinfo{author}{\bibfnamefont{J.-L.} \bibnamefont{Gervais}} \bibnamefont{and}
  \bibinfo{author}{\bibfnamefont{B.}~\bibnamefont{Sakita}},
  \bibinfo{journal}{Phys.Rev.} \textbf{\bibinfo{volume}{D30}},
  \bibinfo{pages}{1795} (\bibinfo{year}{1984}{\natexlab{b}}).

\bibitem[{\citenamefont{Cohen}(1996)}]{Cohen:1996zz}
\bibinfo{author}{\bibfnamefont{T.~D.} \bibnamefont{Cohen}},
  \bibinfo{journal}{Rev.Mod.Phys.} \textbf{\bibinfo{volume}{68}},
  \bibinfo{pages}{599} (\bibinfo{year}{1996}).

\bibitem[{\citenamefont{Kaiser and Leutwyler}(2000)}]{Kaiser:2000gs}
\bibinfo{author}{\bibfnamefont{R.}~\bibnamefont{Kaiser}} \bibnamefont{and}
  \bibinfo{author}{\bibfnamefont{H.}~\bibnamefont{Leutwyler}},
  \bibinfo{journal}{Eur.Phys.J.} \textbf{\bibinfo{volume}{C17}},
  \bibinfo{pages}{623} (\bibinfo{year}{2000}), \eprint{hep-ph/0007101}.

\bibitem[{\citenamefont{Labrenz and Sharpe}(1996)}]{Labrenz:1996jy}
\bibinfo{author}{\bibfnamefont{J.~N.} \bibnamefont{Labrenz}} \bibnamefont{and}
  \bibinfo{author}{\bibfnamefont{S.~R.} \bibnamefont{Sharpe}},
  \bibinfo{journal}{Phys.Rev.} \textbf{\bibinfo{volume}{D54}},
  \bibinfo{pages}{4595} (\bibinfo{year}{1996}), \eprint{hep-lat/9605034}.

\bibitem[{\citenamefont{Sharpe}(1997)}]{Sharpe:1997by}
\bibinfo{author}{\bibfnamefont{S.~R.} \bibnamefont{Sharpe}},
  \bibinfo{journal}{Phys.Rev.} \textbf{\bibinfo{volume}{D56}},
  \bibinfo{pages}{7052} (\bibinfo{year}{1997}), \eprint{hep-lat/9707018}.

\bibitem[{\citenamefont{Chow and Rey}(1997)}]{Chow:1997cu}
\bibinfo{author}{\bibfnamefont{C.-K.} \bibnamefont{Chow}} \bibnamefont{and}
  \bibinfo{author}{\bibfnamefont{S.-J.} \bibnamefont{Rey}}
  (\bibinfo{year}{1997}), \eprint{hep-ph/9712528}.

\bibitem[{\citenamefont{Hasenfratz et~al.}(2007)\citenamefont{Hasenfratz,
  Hoffmann, and Schaefer}}]{Hasenfratz:2007rf}
\bibinfo{author}{\bibfnamefont{A.}~\bibnamefont{Hasenfratz}},
  \bibinfo{author}{\bibfnamefont{R.}~\bibnamefont{Hoffmann}}, \bibnamefont{and}
  \bibinfo{author}{\bibfnamefont{S.}~\bibnamefont{Schaefer}},
  \bibinfo{journal}{JHEP} \textbf{\bibinfo{volume}{0705}}, \bibinfo{pages}{029}
  (\bibinfo{year}{2007}), \eprint{hep-lat/0702028}.

\bibitem[{MIL()}]{MILC}
\urlprefix\url{http://www.physics.utah.edu/~detar/milc/}.

\bibitem[{\citenamefont{Sommer}(1994)}]{Sommer:1993ce}
\bibinfo{author}{\bibfnamefont{R.}~\bibnamefont{Sommer}},
  \bibinfo{journal}{Nucl.Phys.} \textbf{\bibinfo{volume}{B411}},
  \bibinfo{pages}{839} (\bibinfo{year}{1994}), \eprint{hep-lat/9310022}.

\bibitem[{\citenamefont{Bazavov et~al.}(2010)\citenamefont{Bazavov, Toussaint,
  Bernard, Laiho, DeTar et~al.}}]{Bazavov:2009bb}
\bibinfo{author}{\bibfnamefont{A.}~\bibnamefont{Bazavov}},
  \bibinfo{author}{\bibfnamefont{D.}~\bibnamefont{Toussaint}},
  \bibinfo{author}{\bibfnamefont{C.}~\bibnamefont{Bernard}},
  \bibinfo{author}{\bibfnamefont{J.}~\bibnamefont{Laiho}},
  \bibinfo{author}{\bibfnamefont{C.}~\bibnamefont{DeTar}},
  \bibnamefont{et~al.}, \bibinfo{journal}{Rev.Mod.Phys.}
  \textbf{\bibinfo{volume}{82}}, \bibinfo{pages}{1349} (\bibinfo{year}{2010}),
  \eprint{0903.3598}.

\bibitem[{\citenamefont{Aoki et~al.}(2009)}]{Aoki:2008sm}
\bibinfo{author}{\bibfnamefont{S.}~\bibnamefont{Aoki}} \bibnamefont{et~al.}
  (\bibinfo{collaboration}{PACS-CS Collaboration}),
  \bibinfo{journal}{Phys.Rev.} \textbf{\bibinfo{volume}{D79}},
  \bibinfo{pages}{034503} (\bibinfo{year}{2009}), \eprint{0807.1661}.

\bibitem[{\citenamefont{Walker-Loud et~al.}(2009)\citenamefont{Walker-Loud,
  Lin, Richards, Edwards, Engelhardt et~al.}}]{WalkerLoud:2008bp}
\bibinfo{author}{\bibfnamefont{A.}~\bibnamefont{Walker-Loud}},
  \bibinfo{author}{\bibfnamefont{H.-W.} \bibnamefont{Lin}},
  \bibinfo{author}{\bibfnamefont{D.}~\bibnamefont{Richards}},
  \bibinfo{author}{\bibfnamefont{R.}~\bibnamefont{Edwards}},
  \bibinfo{author}{\bibfnamefont{M.}~\bibnamefont{Engelhardt}},
  \bibnamefont{et~al.}, \bibinfo{journal}{Phys.Rev.}
  \textbf{\bibinfo{volume}{D79}}, \bibinfo{pages}{054502}
  (\bibinfo{year}{2009}), \eprint{0806.4549}.

\bibitem[{\citenamefont{Alexandrou et~al.}(2011)\citenamefont{Alexandrou,
  Gregory, Korzec, Koutsou, Negele et~al.}}]{Alexandrou:2011py}
\bibinfo{author}{\bibfnamefont{C.}~\bibnamefont{Alexandrou}},
  \bibinfo{author}{\bibfnamefont{E.~B.} \bibnamefont{Gregory}},
  \bibinfo{author}{\bibfnamefont{T.}~\bibnamefont{Korzec}},
  \bibinfo{author}{\bibfnamefont{G.}~\bibnamefont{Koutsou}},
  \bibinfo{author}{\bibfnamefont{J.~W.} \bibnamefont{Negele}},
  \bibnamefont{et~al.}, \bibinfo{journal}{Phys.Rev.Lett.}
  \textbf{\bibinfo{volume}{107}}, \bibinfo{pages}{141601}
  (\bibinfo{year}{2011}), \eprint{1106.6000}.

\bibitem[{\citenamefont{Witten}(1979)}]{Witten:1979vv}
\bibinfo{author}{\bibfnamefont{E.}~\bibnamefont{Witten}},
  \bibinfo{journal}{Nucl.Phys.} \textbf{\bibinfo{volume}{B156}},
  \bibinfo{pages}{269} (\bibinfo{year}{1979}).

\bibitem[{\citenamefont{Aoki et~al.}(2003)}]{Aoki:2002fd}
\bibinfo{author}{\bibfnamefont{S.}~\bibnamefont{Aoki}} \bibnamefont{et~al.}
  (\bibinfo{collaboration}{CP-PACS Collaboration}),
  \bibinfo{journal}{Phys.Rev.} \textbf{\bibinfo{volume}{D67}},
  \bibinfo{pages}{034503} (\bibinfo{year}{2003}), \eprint{hep-lat/0206009}.

\bibitem[{\citenamefont{Aubin et~al.}(2004)}]{Aubin:2004fs}
\bibinfo{author}{\bibfnamefont{C.}~\bibnamefont{Aubin}} \bibnamefont{et~al.}
  (\bibinfo{collaboration}{MILC Collaboration}), \bibinfo{journal}{Phys.Rev.}
  \textbf{\bibinfo{volume}{D70}}, \bibinfo{pages}{114501}
  (\bibinfo{year}{2004}), \eprint{hep-lat/0407028}.

\bibitem[{\citenamefont{Gasser and Leutwyler}(1984)}]{Gasser:1983yg}
\bibinfo{author}{\bibfnamefont{J.}~\bibnamefont{Gasser}} \bibnamefont{and}
  \bibinfo{author}{\bibfnamefont{H.}~\bibnamefont{Leutwyler}},
  \bibinfo{journal}{Annals Phys.} \textbf{\bibinfo{volume}{158}},
  \bibinfo{pages}{142} (\bibinfo{year}{1984}).

\bibitem[{\citenamefont{Aoki et~al.}(2013)\citenamefont{Aoki, Aoki, Bernard,
  Blum, Colangelo et~al.}}]{Aoki:2013ldr}
\bibinfo{author}{\bibfnamefont{S.}~\bibnamefont{Aoki}},
  \bibinfo{author}{\bibfnamefont{Y.}~\bibnamefont{Aoki}},
  \bibinfo{author}{\bibfnamefont{C.}~\bibnamefont{Bernard}},
  \bibinfo{author}{\bibfnamefont{T.}~\bibnamefont{Blum}},
  \bibinfo{author}{\bibfnamefont{G.}~\bibnamefont{Colangelo}},
  \bibnamefont{et~al.} (\bibinfo{year}{2013}), \eprint{1310.8555}.

\end{thebibliography}

\end{document}